\documentclass[fleqn,usenatbib]{mnras}

\usepackage{newtxtext,newtxmath}

\usepackage[T1]{fontenc}

\DeclareRobustCommand{\VAN}[3]{#2}
\let\VANthebibliography\thebibliography
\def\thebibliography{\DeclareRobustCommand{\VAN}[3]{##3}\VANthebibliography}

\usepackage{graphicx}	
\usepackage{amsmath}	
\usepackage{cancel}
\usepackage[normalem]{ulem}

\usepackage{color}
\usepackage[normalem]{ulem}

\title[Complex Coronal Heating]{On the complex nature of coronal heating}

\author[C.A. Breu et al.]{
C.A. Breu,$^{1}$\thanks{E-mail: cab42@st-andrews.ac.uk (University of St Andrews)}
D.I. Pontin,$^{2}$
E. Priest,$^{1}$
I. De Moortel$^{1,3}$
\\
$^{1}$ School of Mathematics and Statistics, University of St Andrews, St Andrews, Fife KY16 9SS, UK\\
$^{2}$School of Information and Physical Sciences, University of Newcastle, University Drive, Callaghan, 2308, Australia\\
$^{3}$Rosseland Centre for Solar Physics, University of Oslo, PO Box 1029 Blindern, NO-0315 Oslo, Norway\\
}

\date{Accepted XXX. Received YYY; in original form ZZZ}

\pubyear{2015}

\begin{document}
\label{firstpage}
\pagerange{\pageref{firstpage}--\pageref{lastpage}}
\maketitle

\begin{abstract}
A large part of the hot corona consists of magnetically confined, bright plasma loops. These observed loops are in turn structured into bright strands.
We investigate the relationship between magnetic field geometry, plasma properties and bright strands with the help of a 3D resistive MHD simulation of a coronal loop rooted in a self-consistent convection zone layer.
We find that it is impossible to identify a loop as a simple coherent magnetic flux tube that coincides with plasma of nearly uniform temperature and density. The location of bright structures is determined by a complex interplay between heating, cooling and evaporation timescales. Current sheets form preferentially at the interfaces of magnetic flux from different sources. They may also form within bundles of magnetic field lines since motions within magnetic concentrations drive plasma flows on a range of timescales that provide further substructure and can locally enhance magnetic field gradients and thus facilitate magnetic reconnection.
The numerical experiment therefore possesses  aspects of both the flux tube tectonics and flux braiding models. While modelling an observed coronal loop as a cylindrical flux tube is useful to understand the physics of specific heating mechanisms in isolation, it does not describe well the structure of a coronal loop rooted in a self-consistently evolving convection zone.

\end{abstract}




\section{Introduction}

The appearance of much of the magnetically closed solar corona is organized at first sight into  loops of hot and bright plasma apparently confined by magnetic flux tubes that are anchored in the photosphere at both footpoints. While often called a fundamental building block of the solar corona, these loops show substructure on scales of 100 km \citep{2013_Brooks, 2020a_Williams, 2020b_Willams}. What process exactly determines the scale of these bright strands and so a fundamental width across the magnetic field is still subject of active debate. 
In addition, the diffuse coronal emission has been found from Solar Orbiter observations to account for two-thirds of the emission in the quiet Sun \citep{gorman23}.
Proposed coronal heating mechanisms are likely to operate on much smaller length scales than the widths of strands \citep{2013_Brooks}. Current sheets are thought to form on scales of the ion inertial length, which is on the order of a few hundred meters for the corona, while wave dissipation takes place on similarly small scales  \citep[e.g.,][]{priest00}.\\

The term "strand" has to be regarded with some caution, since its meaning depends on the context in which it is used. It can either be linked to the fragmentation of magnetic flux in the solar photosphere and refer to a bundle of magnetic field lines originating from a specific photospheric magnetic concentration, or a plasma strand with approximately uniform density and temperature over a cross-section \citep{2015_Klimchuk}.
While nanoflares and wave dissipation are supposed to occur on scales far below the current resolution of instruments, it appears that loop strands are becoming resolved \citep{2013_Peter,2017_Aschwanden,2021_Williams,2025_Ram}. 
The situation is complicated by the fact that we can neither directly measure the coronal magnetic field nor the temperature and density of the optically thin plasma; all we see is a bright emitting structure in a specific wavelength band. Due to the optically thin nature of the corona, structures along the line of sight overlap, leading to even more uncertainty regarding the 3D structure of coronal loops.
The question of coronal loop substructure is closely intertwined with the coronal heating mechanisms, since it could shed light on the scales at which the mechanisms operate.
A range of physical processes have been proposed to explain the heating of the Sun’s corona to multi-million degree temperatures. These include various wave dissipation mechanisms, as well as the “DC” braiding and coronal tectonics models. Determining the contribution from each mechanism in different parts of the corona has proved challenging, in part due to the necessary simplifications made in numerical simulations. For example, models on active region scales suffer from a poor resolution of individual structures, while most braiding models of local structures are restricted to uniform boundary flux distributions and idealised driving flows. In this paper we describe simulations in which these simplifications are removed by the inclusion of a convection zone at either end of the domain, where the magnetic field is rooted in distinct polarities.
The main difference between different families of models is the timescale of the energy injection at the solar surface compared to the response timescales of the corona. The emergence of magnetic flux from the convection zone into the corona could also provide an important energy contribution \citep{2014_Cheung}.
The first DC models considered braiding of an initially uniform magnetic field by motions at the boundary surfaces \citep{1972_Parker} with heating then taking place in current sheets that form throughout the volume.  
The location of heating in these models is determined by the location of the current sheets. 
A variety of boundary driving profiles and physics within the volume have been considered in modelling efforts -- see the review by \cite{2020LRSP...17....5P}.
Another framework to apply to coronal heating is MHD turbulence \citep{1986_Ballegooijen,1992_Heyvaerts,Rappazzo_2008,Rappazzo_2010}. Here, energy injected by footpoint motions cascades to small scales, where it is dissipated. Indeed, it has been suggested as a prime possibility for heating the diffuse corona \citep{gorman23}.\\ 

One thing that all of the
Parker braiding models have in common is that the two magnetically line-tied boundaries have a uniform distribution of magnetic flux. However, the magnetic field at the solar surface is organised on a large range of spatial scales, from intense flux tubes with kilogauss field strengths and typical sizes of about 100 km in intergranular lanes \citep{2002_Priest} to sunspots. Even the quiet Sun contains intense small-scale flux tubes \citep{2010_Lagg}. 

The braiding model has therefore been  developed by taking into account the discrete distribution of magnetic flux on the solar surface in magnetic concentrations \citep{2002_Priest} to give the flux tube tectonics model, in which the magnetic flux sources are advected by flows on the solar surface. The relative motions of photospheric magnetic flux sources lead to formation and dissipation of current sheets in the corona.
Current sheets form preferentially at the interfaces between magnetic field lines originating from different sources, the so-called separatrices or quasi-separatrix layers. While many simplified models assume a one-to-one correspondence between magnetic flux tubes at each of the footpoints, on the real Sun 
any given magnetic concentration typically connects to many others \citep[e.g.,][]{schrijver2002,2003_Close},
further complicating the picture of coronal heating at the interfaces of flux tubes since clearly separable magnetic flux tubes do not exist.

In the original flux tube tectonics model, photospheric magnetic flux concentrations were considered to be of negligible spatial extent \citep{2002_Priest}. 
This model can therefore be extended to take into account small-scale motions \textit{within} magnetic concentrations that have been found  in abundance in both observations and simulations \citep[e.g.][]{2008_Bonet, 2011_Moll, 2020_Yadav, 2021_Battaglia}. The magnetic field is then sheared by the bulk motions of magnetic flux concentrations while also being braided inside each flux tube by plasma flows inside a magnetic patch. The latter effect is similar to the dynamic braiding picture proposed by \citet{2011_van_Ballegooijen}.
On the real Sun, we are therefore likely dealing with a mixture of the traditional flux tube tectonics and braiding pictures.\\

Motions inside magnetic concentrations often take the form of vortex motions, which can twist flux tubes and lead to the excitation of torsional Alfv\'enic waves. It has also been suggested that vortex motions within intergranular lanes could govern the substructure of coronal loops and determine the strand width \citep{2015_Peter}.\\
Attempts have been made to study the relative contributions of DC and AC heating in complex simulations. \citet{2017_Rempel} calculated the Poynting flux in a simulation of a coronal arcade from magnetic and velocity fields averaged over different time windows, compared it with the instantaneous Poynting flux and found that a significant contribution to the Poynting flux comes from motions on timescales of about an hour. These simulations, however, had a coarse grid spacing of 192 km, thus not properly resolving intergranular lanes and flows within them.\\

In the tectonics model, coronal loop strands would result from the multiple separatrix and quasi-separatrix surfaces embedded in a coronal loop. In models including the response of the plasma to coronal heating, the appearance of coronal loops in emission is determined by the interplay of the temperature and density distribution in the corona and the temperature-dependent contribution function in a specific observed wavelength band \citep{2012_Peter}. Magnetic field and bright strands can have a complex relationship and an observable bright loop does not necessarily correspond to a static or coherent magnetic field structure \citep{2015_Chen,2017_Pontin}. Going one step further, the existence of coronal loops as clearly identifiable compact emitting regions has been called into question \citep{2022_Malanushenko}, while stereoscopic observations indicate that at least some coronal loops seem to be coherent structures with nearly circular cross sections \citep{2025_Ram}. Magnetic field strand sizes measured from numerical simulations are influenced by the grid resolution, with simulations with higher resolutions resolving more fine structure \citep{Breu_2025}.\\

In this study, we investigate how emission, thermodynamic quantities, plasma flows and magnetic connectivity are related and compare to more idealized loop models. 
We use the terms "flux tube", "flux bundle" and "strand", which we define for our simulation as follows. A flux tube consists of a series of magnetic field lines that link one discrete source in or near the photosphere to one other source. A flux bundle, on the other hand, consists of field lines that start from one discrete source and end up at more than one separate source. A strand is a thin flux tube, with approximately uniform temperature and density over its cross section that can be treated as 1D structure for many purposes.
The paper is structured as follows. In section \ref{sec:meth} we describe the numerical setup and analysis methods. In section \ref{sec:results}, we report the main results, with an example of a reconnection event given in section \ref{subsect:recon}. Section \ref{sec:disc} then discusses the findings, followed by the conclusions in section \ref{sec:conc}.

\section{Methods}
\label{sec:meth}

\subsection{Numerical setup}
\label{sec:setup}

We simulate a coronal loop as a straightened-out magnetic flux bundle rooted in two shallow convection zone layers at each loop footpoint. The simulations are run with a version of the 3D radiative MHD code MURaM \citep{2003PhDT_Voegler,2017_Rempel} modified for this kind of setup \citep[see][]{2022_Breu}.\\

The computational box has dimensions of $6\times 6 \times 57$ Mm. The convection layers at each end are 3.5 Mm deep, leading to an effective loop length of 50 Mm. We describe here the results of two separate simulations, undertaken with grid spacings of 60 km and 24 km, respectively. In the following, we will refer to the coronal loop footpoints at each end of the simulation box as FP 1 and FP 2. Since the vertical direction is in our case the direction along the loop axis, we will instead use the coordinate $s$. We consider a time range of ~30 min with an output cadence of ~1 s.\\

The simulations include effects of gravitational stratification, field-aligned Spitzer heat conduction, optically thick grey radiative losses in the photosphere and chromosphere and optically thin losses in the corona.
The chromosphere and corona are assumed to be in local thermodynamic equilibrium (LTE).\\
\subsubsection{Boundary Conditions}

For the bottom boundary, we use open boundaries to mimick the coupling of the simulation domain to deeper layers of the convection zone. All three components of the mass flux and the magnetic field are symmetric with respect to the bottom boundary, so that the gradients of velocity and magnetic
field become zero at the boundary. The gas pressure is decomposed into the mean pressure and a
fluctuation component. The mean pressure component is extrapolated into the ghost cells
assuming a fixed value of pressure taken from the standard solar model at the boundary,
while the fluctuation component is damped. Convection is maintained by specifying the entropy of the inflows at the lower boundary, while it is symmetric in the downflows. The inflow entropy is chosen to achieve a solar-like energy flux within the simulation box. The bottom boundary corresponds to the boundary "Osb" discussed in \citet{2014_Rempel}.
For the initial shallow convection simulation (see below), the top boundary is open for outflows, but closed for inflows. The magnetic field is set to be vertical at the top boundary. The vertical component of the conductive heat flux is set to zero. The horizontal boundaries of the simulation box are periodic.\\

\subsubsection{Coronal extension}

\begin{figure}
\resizebox{\hsize}{!}{\includegraphics{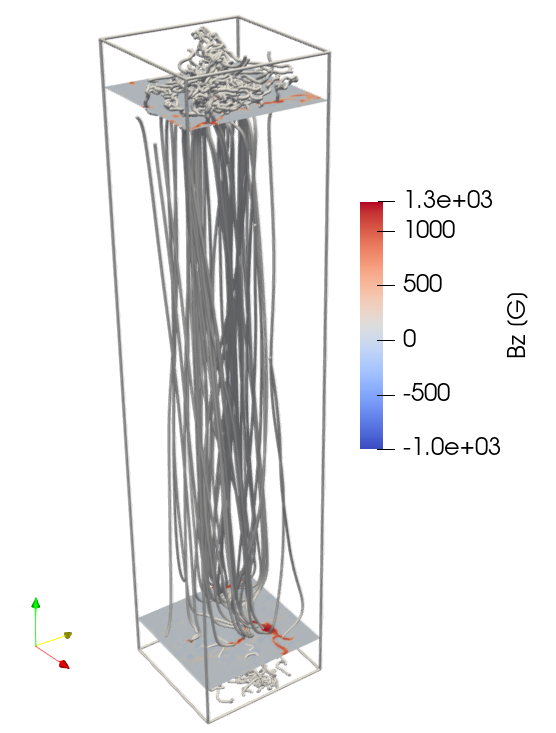}}
  \caption{Simulation setup. Three-dimensional magnetic field lines at t=12.36 min. The slices show maps of the vertical magnetic field component $B_{z}$ at the photosphere at each foodpoint.}
  \label{fig:setup}
\end{figure}

The setup follows the process outlined in \citet{2022_Breu}. We start with a shallow convection zone box without a magnetic field and run the simulation for several hours of solar time until convection has set in, using the standard MURaM code \citep{2003PhDT_Voegler, 2014_Rempel, 2017_Rempel}. We then add a uniform vertical initial field of 60 G and run the simulation for another solar hour until the magnetic field is concentrated in the intergranular downflow lanes. In a next step, the simulation is extended into the corona, assuming a temperature profile following a tanh curve. We then calculate the pressure and density assuming a hydrostatic atmosphere, starting with the mean density and pressure at the top of the initial box. The box is then mirrored about the loop midplane (s=25 Mm) and the mirrored boxes are joined together at the midplane, leading to a total loop length of 50 Mm. Since the magnetic field is vertical at the top boundary of the initial box and is initially the same in the two loop legs, the magnetic field can be easily extended into the corona by assuming a vertical field with the same distribution as at the top of the initial box. The loop is then evolved using the modified MURaM code described in \citet{2022_Breu} as a whole for at least 30 minutes to let the initial transient subside. Due to the buildup of small numerical errors and the inherently chaotic nature of the turbulent convection, the evolution of the magnetic field and flows at the two footpoints diverges with time. The setup is illustrated in Fig.~\ref{fig:setup}.\\
In the following, we will use the low resolution run with a grid spacing of 60 km and a run with a higher grid spacing of 24 km to identify current sheets and study the magnetic field connectivity to ensure the results are not just an artifact of low resolution. The setup of both runs corresponds to the LR (low resolution) and MR (medium resolution) run in \citep{2024_Breu}. Due to the large computational cost and size of the output data for the MR run, we only have a high cadence time series with an output cadence of one second over a time range of $\sim$ 27 minutes for the LR run, while the output cadence for the MR run is 96 s. We therefore use the LR run to study the time-dependent response of the thermodynamic variables to the heating and the field line tracing in time.\\

\subsection{Magnetic connectivity}
\label{sec:magcon}
\begin{figure}
\resizebox{\hsize}{!}{\includegraphics{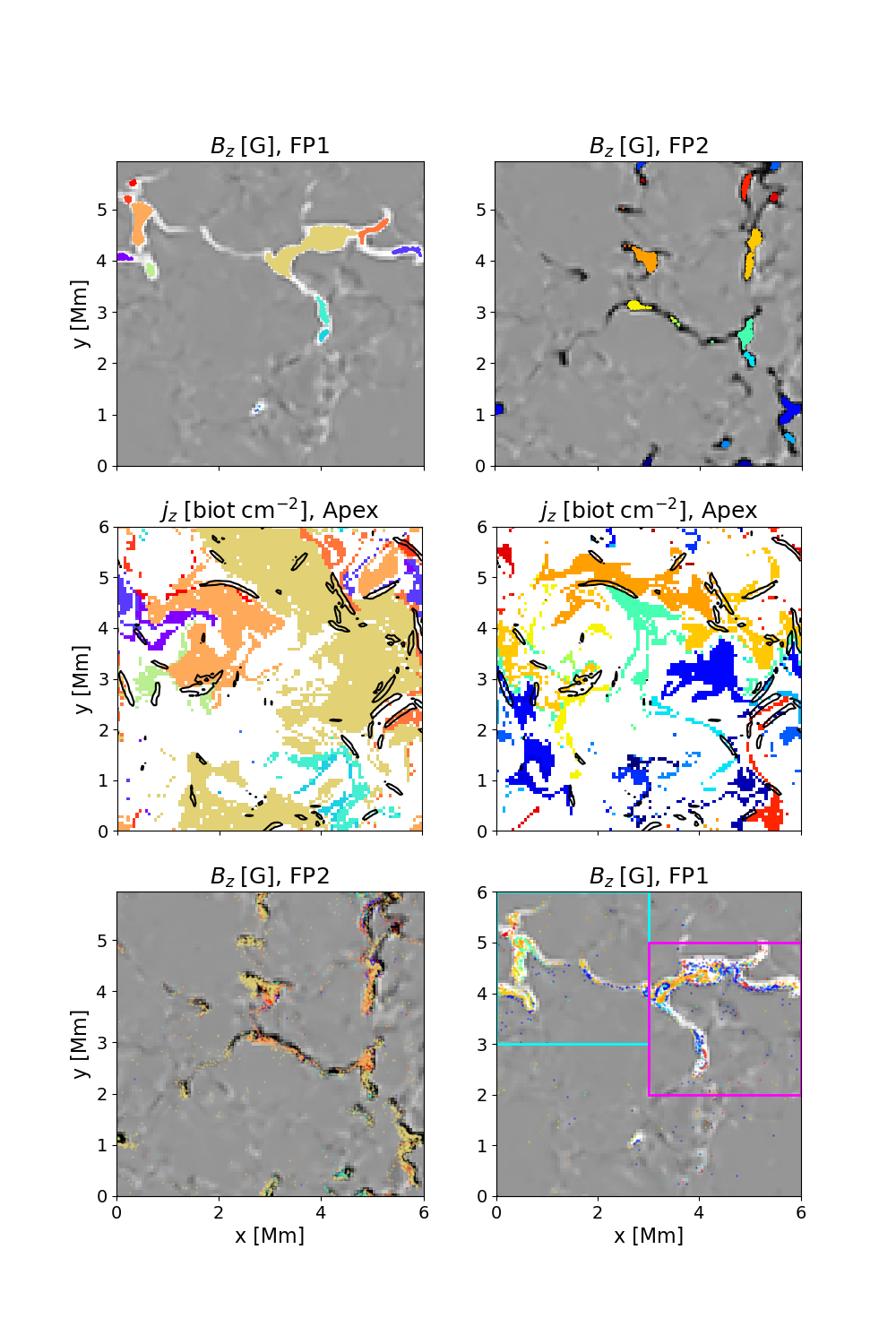}}
  \caption{Flux tube connectivity for run LR at time $t=12.36\; \rm{min}$. Top row: Vertical magnetic field at footpoints 1 and 2. The colored dots show the footpoints of the magnetic field lines traced from the apex that intersect one of the 1 kG magnetic patches. Middle row: Vertical component of the current density at the apex, overlayed with field line seeds colored according to the magnetic concentrations in which they are rooted. The black contours showcase current sheets with current densities two standard deviations above the mean current density. Bottom row: Mapping of magnetic field lines to the opposite loop footpoint. The blue and pink box in the lower right panel mark the positions of the close-ups of the footpoints shown in Fig.~\ref{fig:footpoint_zoom}.}
  \label{fig:connect_fp}
\end{figure}
In the following sections we will discuss the relation of observed structures within our domain to the magnetic fragments at the simulated photosphere. To facilitate this, we
map the magnetic connectivity between the photosphere and corona.
We trace magnetic field lines and generate a map of the connectivity between the coronal loop apex and kilogauss strength magnetic concentrations at the photospheric footpoints following the process outlined in Appendix \ref{appsec:segmentation}. We obtain two mappings, one for each footpoint. The mapping of magnetic field lines between the apex and magnetic concentrations at the footpoints is illustrated in Fig.~\ref{fig:connect_fp}.\\ 
\begin{figure}
\resizebox{\hsize}{!}{\includegraphics{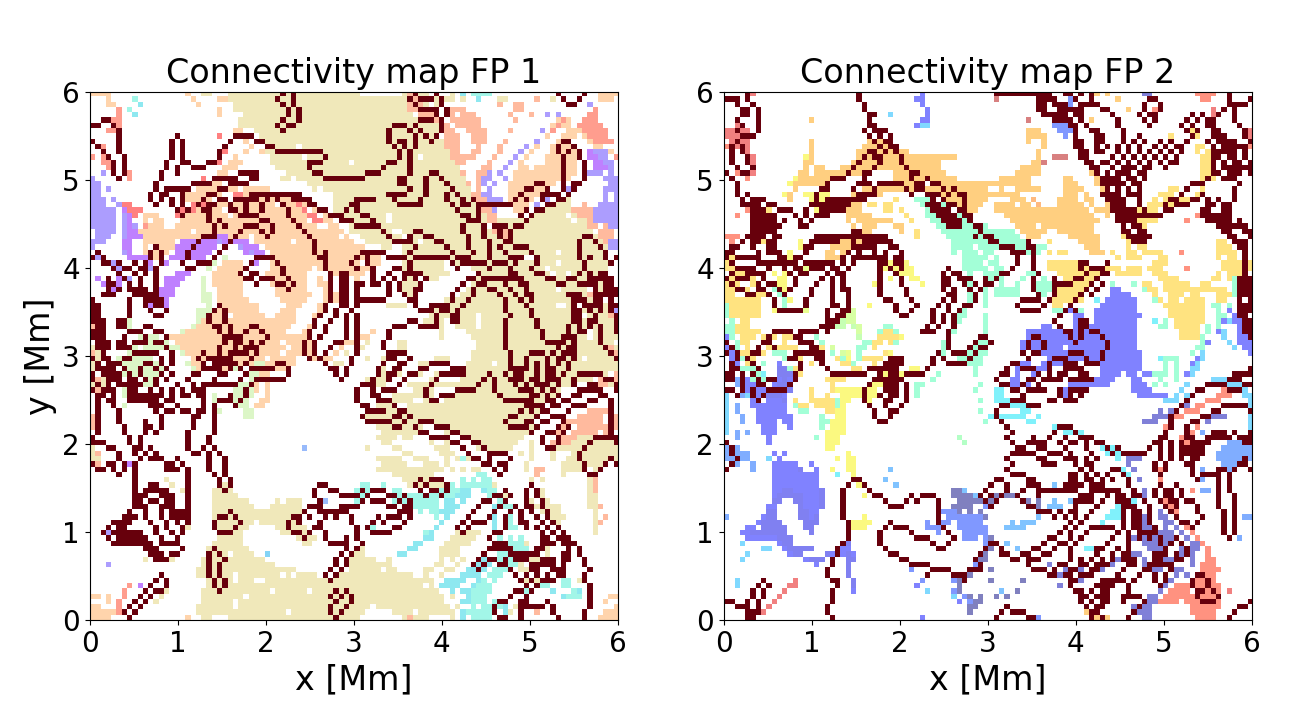}}
  \caption{Fragmentation of magnetic flux bundles in the corona for run LR and the same snapshot as Fig. \ref{fig:connect_fp}. Mapping of field lines from the photosphere to the mid-plane for FP 1 (left) and FP 2 (right). Both panels are overlayed with the edges of the flux bundle maps from the opposite footpoint (dark red contours).}
  \label{fig:octopus}
\end{figure}

One goal of this study is to test whether current sheets form preferentially at the interfaces of flux systems originating from distinct photospheric sources, as predicted by the flux tube tectonics model, or if they are distributed more uniformly throughout the domain as in Parker braiding.
To identify current sheets and check whether a current sheet is located at the boundaries or interfaces of magnetic flux bundles, we develop an automatic procedure, since the manual identification and classification of a large number of small structures for a  large number of simulation snapshots is not feasible. The current sheet categorisation algorithm is described in appendix \ref{appsec:current_auto}. The algorithm first identifies the edges of magnetic flux bundles connected to a specific concentration and then checks for overlap with the current sheets at the apex. A map of all the flux bundle edges computed for both footpoints is shown in Fig.~\ref{fig:octopus}.\\

In addition to tracking field lines in individual snapshots, we also follow field lines in time to study specific events. The field line tracking method and its application to an identified reconnection event are described in appendix \ref{appsec:Btrack} and \ref{appsec:track_recon}.\\
\section{Results}
\label{sec:results}

\subsection{Location of current sheet formation}

\begin{figure}
\resizebox{\hsize}{!}{\includegraphics{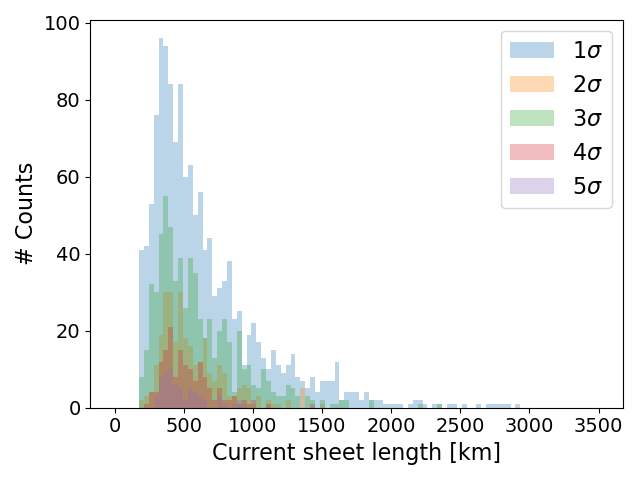}}
  \caption{Distribution of current sheet length for different thresholds imposed on the current density for run LR.}
  \label{fig:current_stats}
\end{figure}

\begin{figure}
\resizebox{\hsize}{!}{\includegraphics{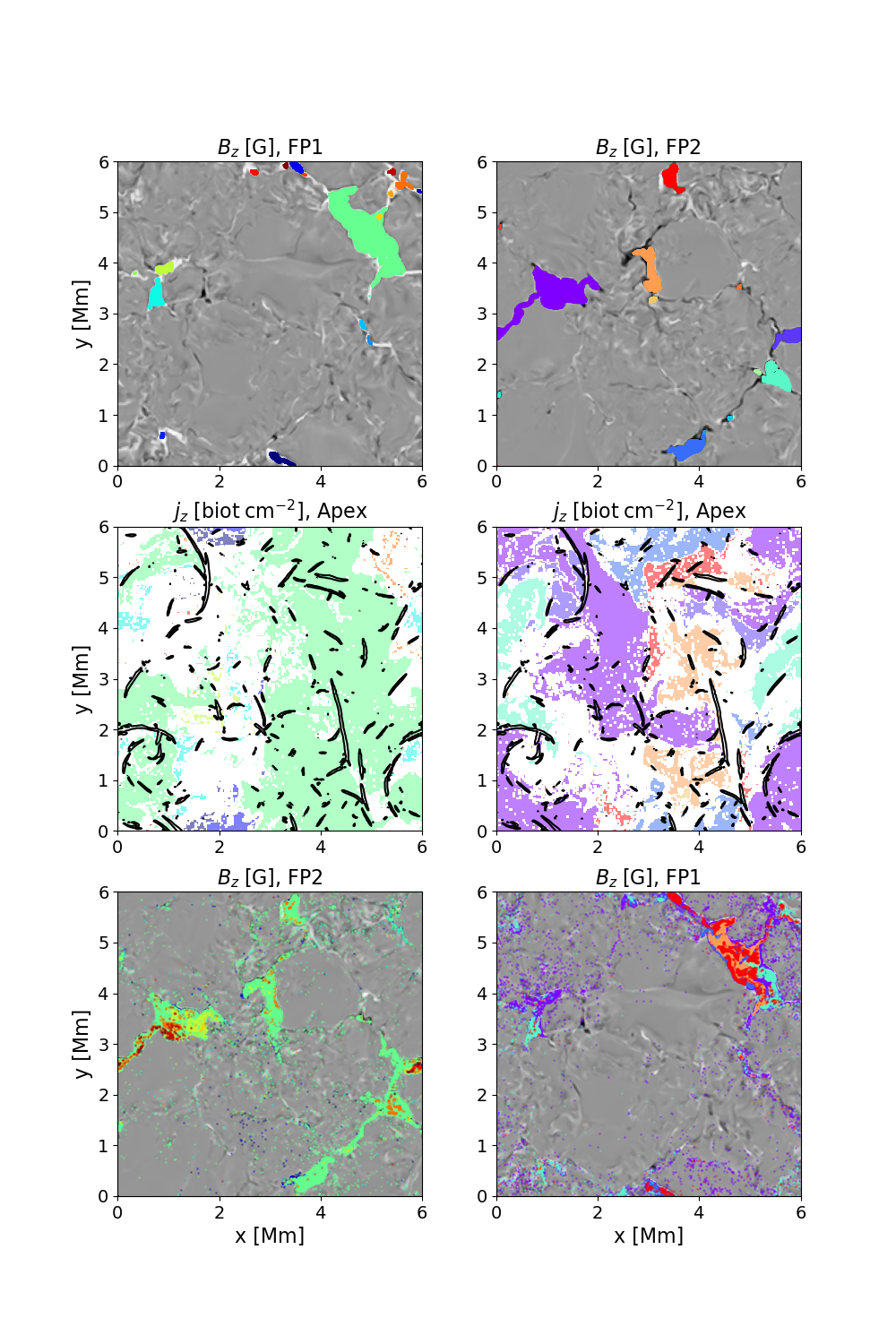}}
  \caption{Flux tube connectivity for a snapshot from simulation run MR with a higher resolution, using a grid size of $\Delta x$= 24 km. Top row: Vertical magnetic field at footpoint 1 and 2. The colored dots show the footpoints of the magnetic field lines traced from the apex. Bottom row: Vertical component of the current density at the apex, overlayed with field line seeds colored according to the magnetic concentrations in which they are rooted. The black contours showcase current sheets with current densities two standard deviations above the mean current density.}
  \label{fig:connect_fp_mr}
\end{figure}

\begin{figure}
\resizebox{\hsize}{!}{\includegraphics{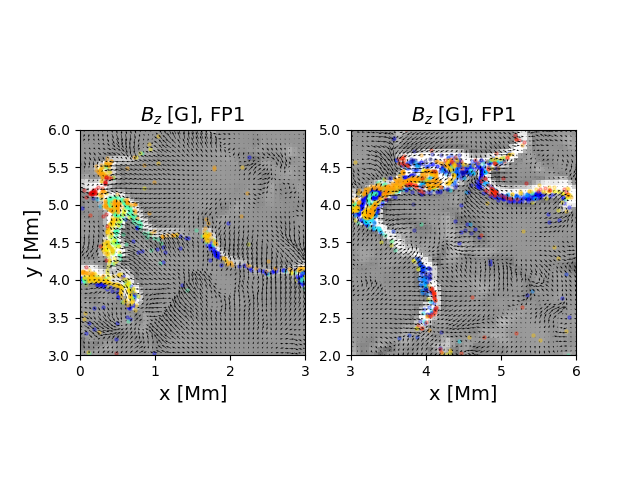}}
  \caption{Close-ups on the lower right panel of Fig. \ref{fig:connect_fp}. The black arrows illustrate the photospheric velocity field. The close-ups correspond to the blue (left) and pink (right) box in Fig.~\ref{fig:connect_fp}.}
  \label{fig:footpoint_zoom}
\end{figure}

Current sheets have a complex three-dimensional geometry and can intersect each other. Due to the difficulty of clearly separating these structures in 3D, we have limited the detection of current sheets in this study to a 2D cut orthogonal to the coronal loop axis, namely in the xy-plane.
Using a threshold on the current density, we find numerous small-scale enhancements of the current density at the loop apex with lengths that are much larger than their widths and which we refer to as current sheets. While many of the structures are very small and only consist of a few grid cells, a few current sheets have lengths of several Mm within the horizontal plane at the loop apex. Automatically determining the length of the current sheets is a nontrivial task due to the curvature and complex shape of the current concentrations. We make use of the regionprops function of the scikit-image python package. The current sheet length is determined by fitting an ellipse to the detected connected region with the same normalized second central moments as the selected region. Structures consisting of less than five connected grid cells are not considered. We then take the major axis of the ellipse as a proxy for the current sheet length. A distribution of the lengths of the current sheets is shown in Fig.~\ref{fig:current_stats}. It is clear that small-scale structures are the most numerous, with the distribution showing a peak at a length of about 500 km, whilst falling off for longer current sheets. The number of current sheets and the maximum length decrease with increased threshold used for current sheet detection.

\begin{table}
	\centering
	\caption{Percentage of current sheets overlapping with flux bundle edges depending on the current density threshold $t_{cs}$ (columns) and the minimum overlap between broadened current sheets and flux bundle edges (rows) for the LR run. The upper row for each overlap refers to the FP 1 mapping and the bottom row to the FP 2 mapping.}
	\label{tab:overlap_lr}
	\begin{tabular}{lccccc} 
		\hline
		   & 1 $\sigma$ & 2 $\sigma$ & 3 $\sigma$ & 4 $\sigma$ & 5$\sigma$ \\
		\hline
		1 pix & 84.7 & 80.6 & 78.8 & 71.1 & 46.7\\
              & 78.8 & 77.9 & 76.8 & 70.9 & 44.1\\
		5 pix & 67.1 & 64.7 & 64.7 & 57.7 & 34.9\\
              & 64.8     &   62.2    & 58   &  55.2 & 37.3\\
		adapt. 20 \% & 44.3 & 55.2 & 58.8 & 57.3 & 40.4\\
                     &  39.7     &  48.8     & 54.9 & 54.1 & 37.4\\
        adapt. 10 \% & 73.3 & 72 & 72.2 & 65.8 & 44.1\\
                     & 65.6 & 70.2 & 70.3 & 65.5  & 44.1\\
		\hline
	\end{tabular}
\end{table}

\begin{table}
	\centering
	\caption{Percentage of current sheets overlapping with flux bundle edges depending on the current density threshold $t_{cs}$ (columns) and the minimum overlap between current sheets and flux bundle edges (rows) for the MR run. The upper row for each overlap refers to the FP 1 mapping and the bottom row to the FP 2 mapping.}
	\label{tab:runs}
	\begin{tabular}{lccccc} 
		\hline
		   & 1 $\sigma$ & 2 $\sigma$ & 3 $\sigma$ & 4 $\sigma$ & 5$\sigma$ \\
		\hline
		1 pix & 64.1 & 62.1  & 62.9 & 66.7 & 68.2\\
              & 72.5  & 76.8 & 83.8 & 88.9 & 90.9\\
		5 pix & 54.2 & 45.3 & 54.8 & 58.3 & 63.6\\
              & 66 & 65.3 & 64.5 & 66.7 & 72.7\\ 
        10 pix & 32.7 &  30.5    &   27.4   &  33.3    & 31.8\\
              & 43.5  &  35.8    &   38.7   &   38.8   & 36.4\\   
		adapt. 20 \% & 33.3 & 33.7    &         45.2       &     47.2      &  54.5  \\
                     & 41.8 & 47.4 & 56.5 & 58.3 & 63.6\\
        adapt. 10 \% & 56.2 & 49.5 & 53.2 & 55.6 & 63.6\\
                     & 64.7 & 69.5 & 74.2 & 80.6 & 86.4 \\
		\hline
	\end{tabular}
\end{table}

If the current sheet formation results from relative motion between magnetic flux originating from different sources, the current sheets should be mostly aligned with the edges of bundles of magnetic field lines originating from a photospheric magnetic patch.
We use our algorithm described in appendix \ref{appsec:current_auto} to detect edges of magnetic flux bundles and determine whether the current sheets overlap with magnetic flux bundle edges. 
While the method checks for overlapping structures, it cannot directly measure how well aligned a current sheet is with the edges. A current sheet crossing a flux bundle edge at a right angle would still be classified as an "edge" current sheet. We therefore set a threshold on the minimum number of grid cells the current sheet is required to overlap with the bundle edges. We experiment with various thresholds on the current density and on the number of overlapping grid cells between a detected current sheet and bundle edges. Since the current sheets can have vastly different sizes, we also test an adaptive threshold for the overlap requiring a certain percentage of the area of the current sheets to overlap with edges instead of a fixed number of grid cells. We average our result over 25 snapshots taken over a simulation runtime of 27 minutes. The results for minimum overlaps of one and five grid cells (300 km) as well as adaptive thresholds of 10\% and 20\% of the total area of the current sheet are summarized in Table \ref{tab:overlap_lr}. Not every strong current sheet forms at the edge of a flux bundle, and not every interface between discrete bundles gives rise to strong current sheets. The percentage of detected "edge" current sheets depends strongly on the required overlap between the current sheets and edges. For a fixed threshold, the percentage is higher for lower thresholds on the current density, leading to more numerous and larger current sheets, and a smaller threshold on the required number of overlapping grid cells. If at least five aligned grid cells are required (corresponding to a length of about 300 km), about two-thirds of detected current sheets are aligned with flux bundle edges and interfaces for thresholds on the current density of up to $\mu+2\sigma$. This number is similar for an adaptive threshold of 10\% of grid cells, while it drops to about half for a required overlap of 20\%.

To test whether our result is an artifact of resolution, we repeat the current sheet identification with simulation cubes with a higher resolution of 24 km.
For higher numerical resolution, the internal structure of magnetic concentrations as well as the cascade to small scales in the atmosphere is better resolved. We would therefore expect a weaker association between current sheets and edges, with instead more current sheets driven by internal braiding. 
 The result is shown in Fig.~\ref{fig:connect_fp_mr}. 
The percentage of edge current sheets (or "tectonics current sheets") indeed drops to about two thirds without a threshold on the overlap and even further with a threshold. Interestingly, the percentage of edge current sheets increases with the detection threshold imposed on the current density. This suggests that a majority of the strongest current sheets do preferentially form at interfaces of magnetic field structures coming from different flux sources, while the "braiding" current sheets tend to be weaker.  There is quite a large difference between the mapping from each footpoint, with a lower percentage of edge current sheets for FP 1. For that footpoint, most of the magnetic flux at the apex is connected to the large magnetic concentration in the right upper quadrant, while the mapping for FP 2 is more fragmented. This indicates that on the Sun, the balance of energy release between tectonics (edge current sheets) and braiding (internal current sheets) may also vary between regions, e.g., Quiet Sun, active region or plage.\\

The total volumetric heating rate $Q_{tot}$ includes resistive and viscous heating. The heating rates are calculated from the diffusive numerical fluxes. The method is described in detail in \citet{2017_Rempel}. The total heating rate integrated over the masks covering the flux bundle edges at the apex is about 28\% of the total heating rate integrated over the loop cross section, while the masks outlining the edges cover about 35\% of the loop cross section. The heating is thus below the average heating rate. This is due to the current sheets being strongly intermittent and only enhanced along part of the flux bundle edges. 
The concept of a magnetic flux bundle by itself is problematic, since we do not find well-defined magnetic flux tubes connecting one magnetic patch on one footpoint to another at the other footpoint. Instead, magnetic field lines anchored in one flux patch connect to multiple flux patches at the other footpoint, while also being braided around each other. A zoom-in on the distribution of magnetic field lines from different magnetic flux patches originating from the respective opposite footpoint within a magnetic flux concentration is shown in Fig. \ref{fig:footpoint_zoom}. Not only is each magnetic patch connected to several other patches, but the field lines connected to different patches do not originate from compact, nearly circular sources, but from elongated structures intertwined with other field lines. 

\subsection{Relation between Magnetic Strands and Plasma Strands}

\begin{figure}
\resizebox{\hsize}{!}{\includegraphics{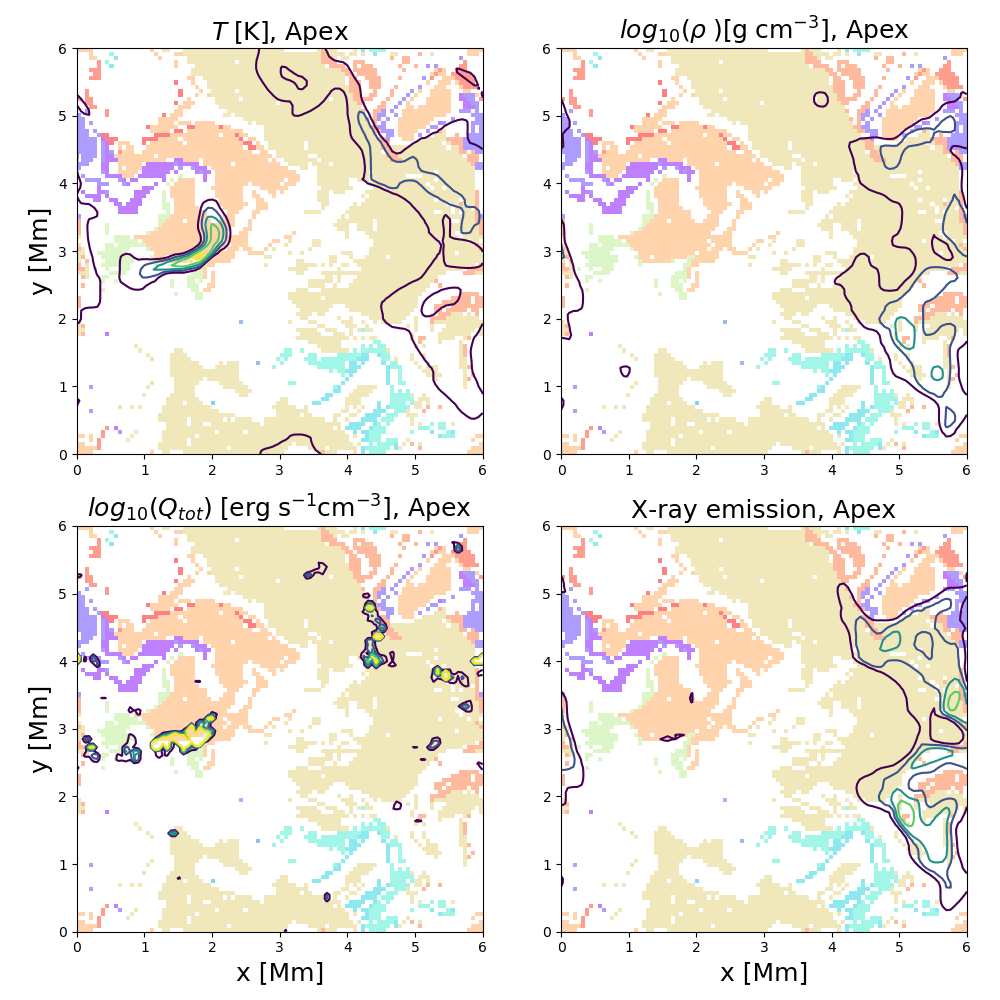}}
  \caption{Relation between various quantities and flux tube connectivity for run LR and the same snapshot as Fig. \ref{fig:connect_fp}. The contours show levels of various thermodynamic quantities and the X-ray emission from one (dark blue) to five (yellow) standard deviations above the mean. Top row: Temperature and density at the loop apex. Bottom row: Total heating comprised of viscous and resistive heating (left) and X-ray emission at the loop apex (right). The plots are overlayed with field line seeds colored
according to the magnetic concentrations in which the field lines are rooted. The footpoint colorings are the same as in the left column of Fig.~\ref{fig:connect_fp}. The contours outline areas where the respective quantities are elevated by 1-5 standard deviations above the mean.}
  \label{fig:emissivity}
\end{figure}

\begin{figure}
\resizebox{\hsize}{!}{\includegraphics{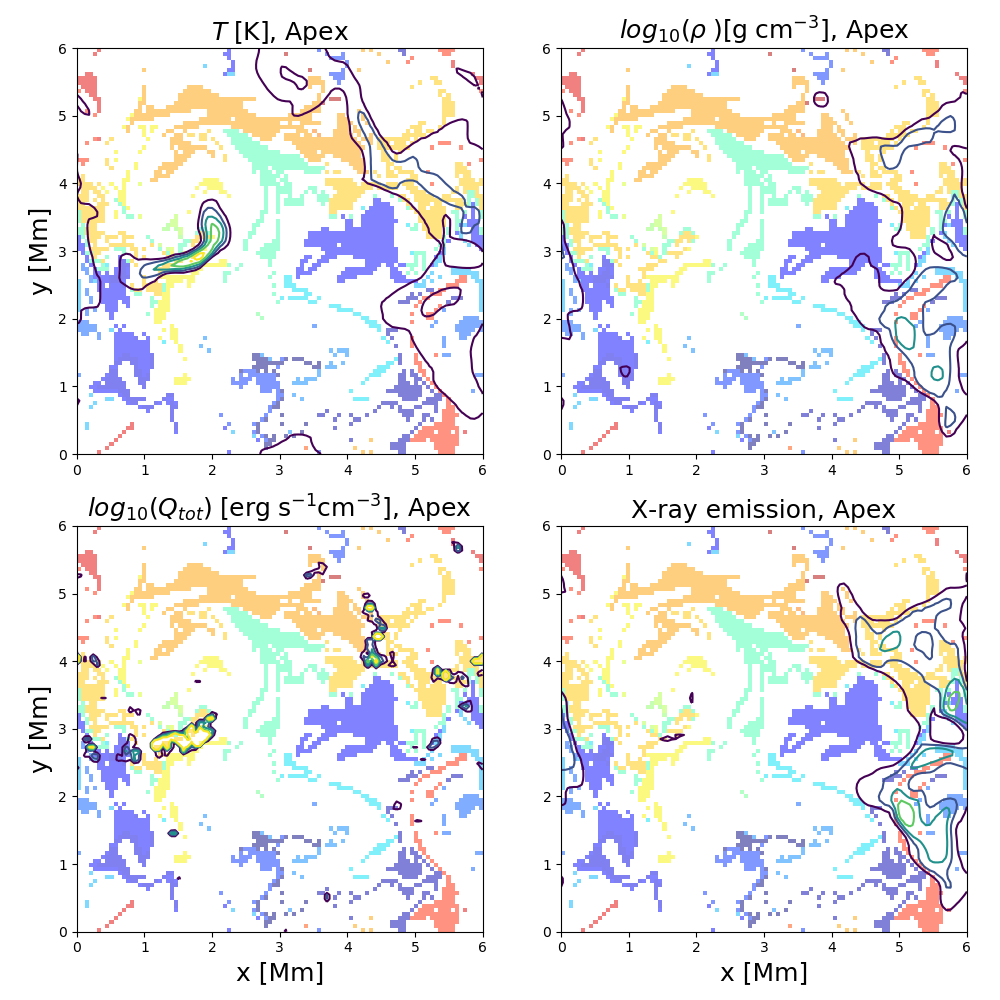}}
  \caption{Relation between various quantities and flux tube connectivity for footpoint FP 2 for run LR and the same snapshot as Fig. \ref{fig:connect_fp}. Top row: Temperature and density at the loop apex. Bottom row: Total heating comprised of viscous and resistive heating (left) and X-ray emission at the loop apex (right). The contours show levels of various thermodynamic quantities and the X-ray emission from one to five standard deviations above the mean. The plots are overlayed with field line seeds colored
according to the magnetic concentrations in which the field lines are rooted. The footpoint colorings are the same as in the right column of Fig.~\ref{fig:connect_fp}. The contours outline areas where the respective quantities are elevated by 1-5 standard deviations above the mean.}
  \label{fig:emissivity2}
\end{figure}

\begin{figure*}
\resizebox{\hsize}{!}{\includegraphics{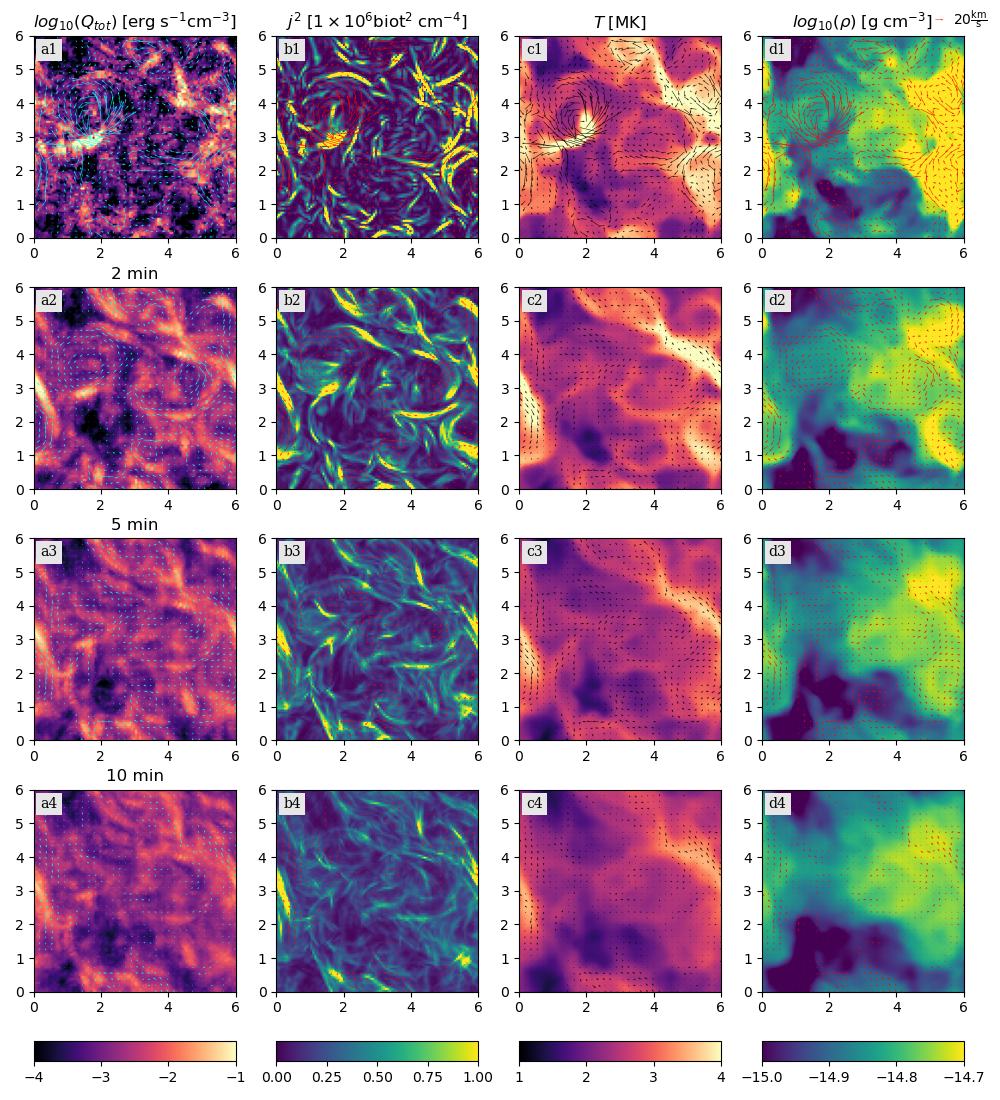}}
  \caption{Time-averaged quantities at the loop apex for run LR. From left to right: Volumetric heating rate (a) comprised of the sum of the viscous and resistive heating rate, squared current density (b), temperature (c) and mass density (d). The top row shows the instantaneous quantities, while the second to last row shows the same quantities averaged over consecutively large time windows, from 2 minutes to 10 minutes preceding the reconnection event at 12.36 min.}
  \label{fig:time_avg}
\end{figure*}

We calculate synthetic emission to test the correspondence between bright plasma strands and magnetic structures. The X-ray emission is computed using the temperature-dependent response function $K(T)$ of the Al-
Poly filter of the X-Ray Telescope on board the Hinode spacecraft (XRT; \citet{2007_Golub}). The synthetic intensity measured by XRT for a grid cell is then given by $\varepsilon = K(T)n_{e}^{2}$, where $n_{e}$ is the electron number density. The response function $K(T)$ is given in \citet{2007_Golub}.\\

There is no clear correspondence of magnetic flux bundles connected to a specific magnetic concentration and bright strands. The relation between heating rate, temperature, density and X-ray emission and field line mapping from both photospheres is shown in Figs. \ref{fig:emissivity} and \ref{fig:emissivity2}. The strongest heating occurs at the edge of a flux bundle in the upper right quadrant of the simulation box cross-section. Temperature and mass density show a less clear picture, with the density in particular reaching the highest values outside  any of the flux bundles connected to the strongest magnetic patches. The X-ray emission mainly follows the distribution of the density due to the quadratic dependence of optically thin losses on the electron density. The X-ray emission therefore partially reaches its highest value outside regions connected to kG concentrations.

While heating rate, current density and the temperature peaks are concentrated in a few narrow sheets, the time-averaged quantities have a much smoother distribution. Thermodynamic quantities at the loop apex averaged over several time windows are shown in Fig. \ref{fig:time_avg}. The density (column d) is highest in areas which are consistently heated over several minutes. The X-ray emission is therefore dependent on the history of the heating and cooling.
While instantaneous heating events are small and occur on very short timescales, heating events cluster in both space and time. This, as well as the slower timescales of density evolution, lead to an overall slower evolution of the temperature and density distribution than the evolution of local heating events. Additionally, the delayed response of the heating leads to bright strands larger than individual heating events. The fraction of the cross-sectional area with increased density is larger than the instantaneous fraction filled with heating events. Since the emission is more strongly dependent on density than on temperature, this broadens the emitting regions. While the heating event at [x,y]=[2,3] Mm is very prominent at time 12.36 min (panel a1 of Fig.~\ref{fig:time_avg}), it is not one of the most consistently heated regions. For example, the upper right quadrant of the simulation box cross-section has a larger  heating rate when averaging over a 10 minute window (panel a4) and is one of the densest and brightest regions (see column d of Fig.~\ref{fig:time_avg} and the bottom right panels of Fig.~\ref{fig:emissivity} and Fig.~\ref{fig:emissivity2}).\\
An order of magnitude estimate for the conductive and radiative cooling timescales based on \cite{1994_Cargill} is given by:
\begin{align}
    \tau_{c}&=4\times 10^{-10}\frac{nL_{half}^{2}}{T^{5/2}}\rm{s},\\
    \tau_{r}&=\frac{3k_{B}T^{1-\alpha}}{n\chi}\rm{s}
\end{align}
where $L_{half}$ is the loop half length, $n$ is the electron density, $T$ is the temperature, and $\alpha$ and $\chi$ are parameters in a fit to the  optically thin radiative loss function of the form $P_{rad}=\chi T^{\alpha}$.
Another time scale of interest is the compression/expansion timescale $\tau_{cmp}$, which is defined as follows:
\begin{equation}
    \tau_{cmp} = \frac{e_{int}}{|(e_{int}+P)\nabla\cdot \mathbf{v}|} \rm{s},
\end{equation}
where $e_{int}$ is the internal energy density, P is the plasma pressure and $v$ is the velocity.
The mean temperature in the loop mid-plane for the snapshot shown in Fig.~\ref{fig:connect_fp} is 2.81 MK and the mean electron density is $9.48\times 10^{8}\; \rm{cm^{-3}}$.
The loop half-length is 25 Mm. The estimate for the conductive cooling time is then 179 s. For the given mean apex temperature the fit used in \citet{1994_Cargill} is $P_{rad}=3.5\times 10^{-7}\cdot T^{-2.5}$. The radiative cooling timescale is then $4.64\times 10^{4}$ s, which is much larger than timescales of interest. The radiative cooling can therefore be neglected compared to cooling by heat conduction. Coronal loops fill with plasma in response to heating events by chromospheric evaporation. The axial speed in the corona varies strongly in time and over the cross-section of the simulation box. The mean upflow velocity at a height of 5 Mm in the low corona is 10-20 $\rm{km\; s^{-1}}$, leading to evaporation timescales of up to half an hour. Locally, however, upflow speeds can be much larger, with speeds of 70-80 $\rm{km\; s^{-1}}$ present in patches and peak speeds in the corona of up to 140 $\rm{km\; s^{-1}}$. The evaporation timescale then becomes comparable to the conduction timescale. \citet{2022_Breu} and \citet{2025_Johnston} show that the density increase on heated strands sets in shortly after the temperature increase. While the evaporation timescale for high-speed upflow regions is compatible with observations which report loop brightening within minutes \citep{chitta20a}, care has to be taken regarding the interpretation of the mechanism behind the brightening. Since both timescales are comparable, it is not clear whether an observed brightening or darkening is due to evaporation of material or shifts of the plasma temperature towards or away from the peak formation temperature of a specific line or wavelength range. The expansion/contraction time scale varies strongly over the loop cross section, from about 5 s in the regions with strongest heating to $10^{6}$ s, with an average of around 900 s. The average time scale is larger than the estimates for the evaporation and conductive cooling, but locally the expansion or contraction of the plasma can have a strong influence on the temperature evolution. Here we are just making simple order-of-magnitude estimates of the timescales. For a fuller analysis of the relevant timescales for a coronal magnetic reconnection event, see \citet{2025_Sen}. 
\\

In the study by \citet{2025_Howson}, the current distribution differs between a tectonics model of driven flux tubes and braiding in uniform field. In the tectonics case, the time-averaged current density forms a honeycomb pattern outlining the edges of the tectonics flux tubes, while having a more diffuse distribution for the braiding setup. In our simulation, the averaged current density is distributed more evenly over the cross-section of the simulation box, suggesting an extra contribution from internal braiding within the flux tube.\\
While field lines adjacent to each other at flux tube edges might have a wide separation at the footpoints and be subject to different driving motions leading to shearing of the field, this is true more generally in our case due to the complex connectivity of the magnetic field. Even field lines belonging to magnetic field coming from one magnetic patch at one footpoint might be connected to multiple concentrations at the other footpoint, leading to internal shear within the bundle in addition to the internal braiding caused by flows within magnetic patches.  
This combination of a complex field line topology that is continuously rearranged and the delayed nature of the appearance of emitting plasma relative to the heating breaks the correspondence between the actual (3D) field structure and observed loops. While the magnetic field is still frozen into the plasma due to the frozen flux theorem, coherent magnetic field structures do not necessarily correspond to contiguous bright plasma structures. This is complicated further by the line-of-sight integration of the optically thin plasma.

\subsection{Relation between Magnetic Strands and Flows}

\begin{figure*}
\resizebox{\hsize}{!}{\includegraphics{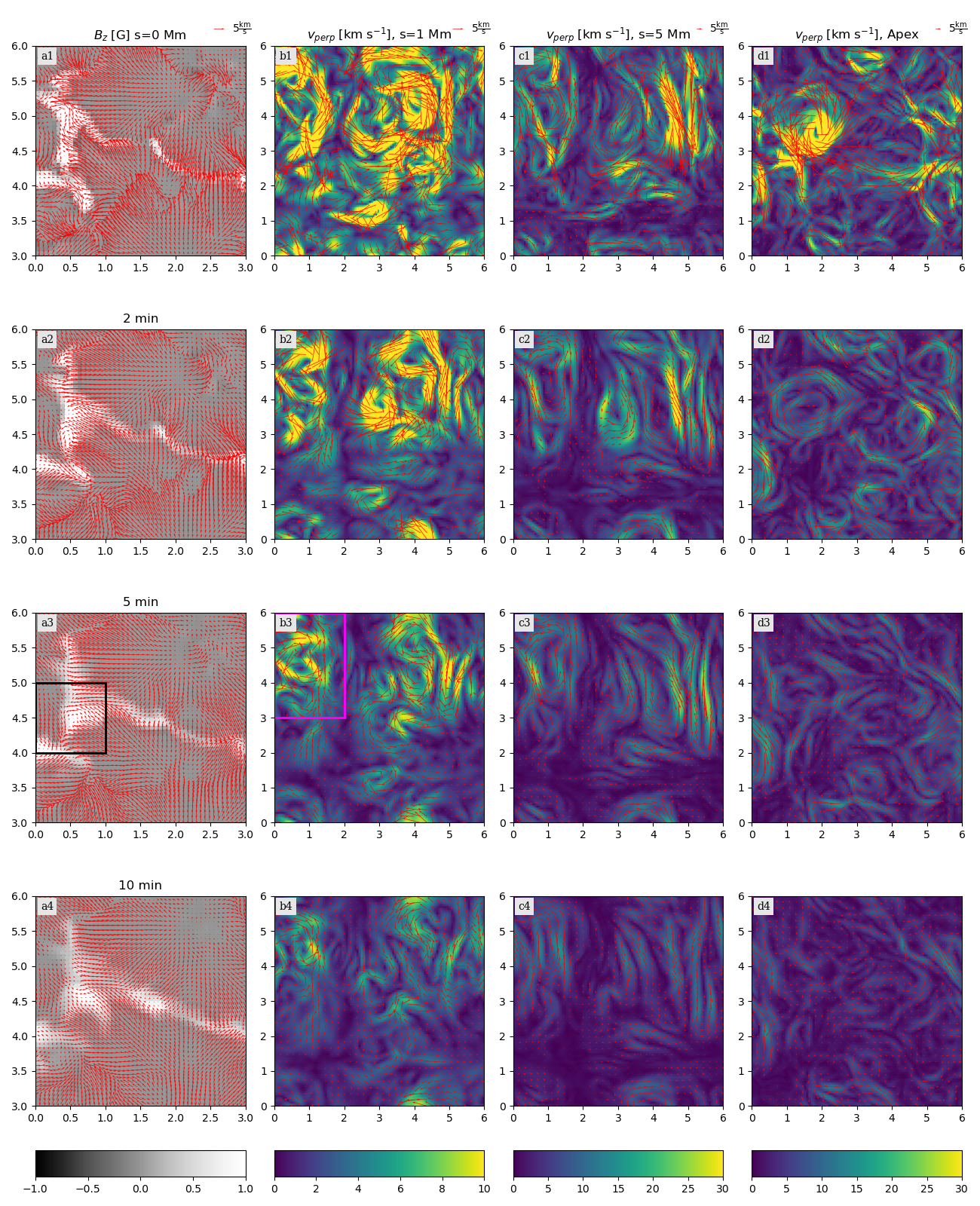}}
  \caption{Time-averaged magnetic field and velocity field at different heights for run LR. The leftmost column (a) shows the photospheric magnetic field for different averaging windows, from the instantaneous field at 40 min, to 2 minutes, 5 minutes and finally 10 minutes averaging time.  Columns b-d: Velocity field for different averaging time windows and heights. The time average was performed over 2-10 minutes up to the reconnection event at 12.36 min. The red arrows illustrate the direction of the velocity field. The boxes highlight a persistent swirling structure with a lifetime of $\geq$ 5 min in an intergranular lane (black box) and in the chromosphere at a height of 1 Mm (pink box). }
  \label{fig:vtavg}
\end{figure*}

Internal twisting within photospheric magnetic flux concentrations could lead to further substructure in coronal loops \citep{2015_Peter}.
Similar to earlier photospheric, chromospheric and coronal simulations and observations, at each timestep multiple rotating plasma structures can be found in photospheric intergranular lanes, the chromosphere and corona. Chromospheric swirls are considered a potential mechanism for launching  torsional Alfv\'en waves into the corona. Although most swirls have short lifetimes  \citep{2018_Giagkiozis, 2023_Tziotziou}, longer-lived structures have been observed, up to a timescale of hours \citep{2018_Tziotziou}. In a coronal hole environment with open magnetic field, the twist induced by photospheric motions would propagate away without altering the magnetic structure for an extended amount of time or increasing the magnetic energy of the configuration. On the other hand, in a magnetically closed structure, the coronal response would depend on the relation between the timescales of photospheric flows and the Alfv\'en crossing time. If the Alfv\'en crossing time was much shorter than the lifetime of a photospheric vortex, the swirling motion would lead to the formation of a twisted flux tubes instead of launching a short-lived torsional perturbation. Such motions have been suggested by Parker in the context of the braiding scenario \citep{1982_Parker,1983_Parker}.

The loop cross-section at different heights shows a velocity field with a complex structure of vortex and shear flows. The strongest heating event at time 40 min occurs at the boundary of a prominent vortex flow. We search for long-lived vortex flows by averaging the $v_{x}$ and $v_{y}$ components of the velocity. Computing the magnitude of the perpendicular velocity from these time-averaged components leads to the disappearance of short-lived high-amplitude structures. The magnetic field at the photosphere and the velocity field at different heights for averaging time windows from 2 min to 10 min are plotted in Fig. \ref{fig:vtavg}.

While the loop apex does not contain discernible large-scale structures living longer than about two minutes over the examined time range of ten minutes, the vortex flow in the upper left quadrant of panels b1-d1 is still discernible in the flow field averaged over ten minutes at lower heights of one Mm and five Mm above the solar photosphere (see panels b4 and c4). The rotational motion is also discernible inside the magnetic concentration forming the footpoint of the vortex for averaging time scales less than five minutes (see the black box in panel a3 in Fig.~\ref{fig:vtavg}). The pink box in panel b3 outlines the long-lived vortex flow in the chromosphere above the photospheric vortex.\\

In order to test whether a DC or an AC mechanism is in operation, we need to compare the timescales of the driving velocities to the Alfv\'en travel time through the loop.
The Alfv\'en travel time is computed as
\begin{equation}
    \tau_{A}(s)=\int_{0}^{s} \frac{ds}{\langle v_{A}\rangle},
\end{equation}
where $s$ is the coordinate along the loop axis and $\langle v_{A}\rangle =\langle |B|/\sqrt{4\pi\rho}\rangle$ is the mean Alfv\'en speed  averaged over the loop cross-section. The travel time is large in the chromospheric part of the simulation domain due to the high density and consequentially lower Alfv\'en speed but small in the coronal part. Overall, the total travel time from one footpoint to the apex is just over five minutes. A chromospheric vortex flow persisting over ten minutes should therefore be able to lead to the creation of a twisted flux rope.\\ 

All MHD simulations of the solar corona suffer from the problem of unrealistic transport coefficients, leading to resistivities that are much too high compared to the real Sun. The magnetic Prandtl number $P_{m}=\frac{\nu}{\eta}$ determines the ratio of viscosity $\nu$ and resistivity $\eta$.
Due to the high Prandtl number of the plasma in the solar corona, energy dissipation occurs mostly through viscosity and resistive heating is expected to be insignificant. Instead, heating should occur through the thermalization of flows driven by the Lorentz force \citep{2017_Rempel}. Due to the high numerical resistivity in numerical codes, the Prandtl number in simulations is limited to significantly lower values than in the real solar atmosphere.
The MURaM code has an adjustable parameter that determines the diffusivity of the numerical scheme. Since this parameter can be set to different values for the magnetic field and the velocity field, it can be used to modify the effective Prandtl number \citep{2014_Rempel,2017_Rempel}, depending on whether the effective numerical resistivity or viscosity is higher. Whilst it is not possible to reach the high Prandtl numbers expected in the solar atmosphere, we run the simulations in a high Prandtl number setting 
so that the viscous heating dominates over the resistive heating.

\begin{figure}
\resizebox{\hsize}{!}{\includegraphics{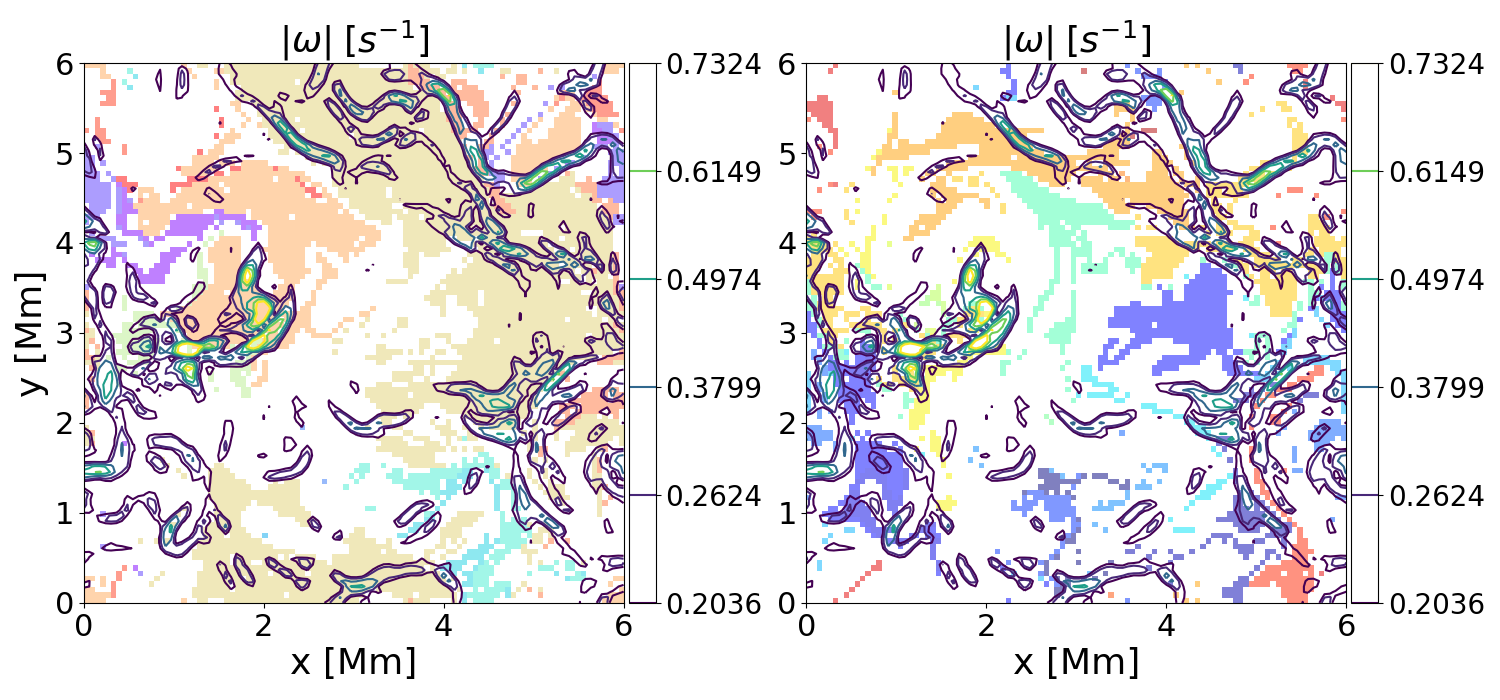}}
  \caption{Relation between magnetic flux bundles and the magnitude of vorticity (unfilled contours) for run LR for the same snapshot as Fig.~\ref{fig:connect_fp}. The colored patches show the mapping of the magnetic field at the apex to different magnetic concentrations at the footpoints for FP 1 (left) and FP 2 (right). The contours show levels of vorticity from one (dark blue) to five standard deviations (yellow) above the mean.}
  \label{fig:vort}
\end{figure}
Small scales in the velocity field, such as small-scale vortices, form dissipative structures where viscous heating occurs. The magnitude of the vorticity $\mathbf{\omega}=|\nabla\times \mathbf{v}|$ can then be used to identify these structures.
The vorticity is enhanced along some of the edges of the flux bundles identified by mapping the magnetic field from the photosphere to the apex. Not all flux bundle interfaces, however, are associated with enhanced vorticity. Instead of outlining the interfaces of flux tubes as in \citet{2025_Howson}, the vorticity is concentrated in a few smaller, sheet-like regions as shown in Fig. \ref{fig:vort}.

\begin{figure*}
\resizebox{\hsize}{!}{\includegraphics{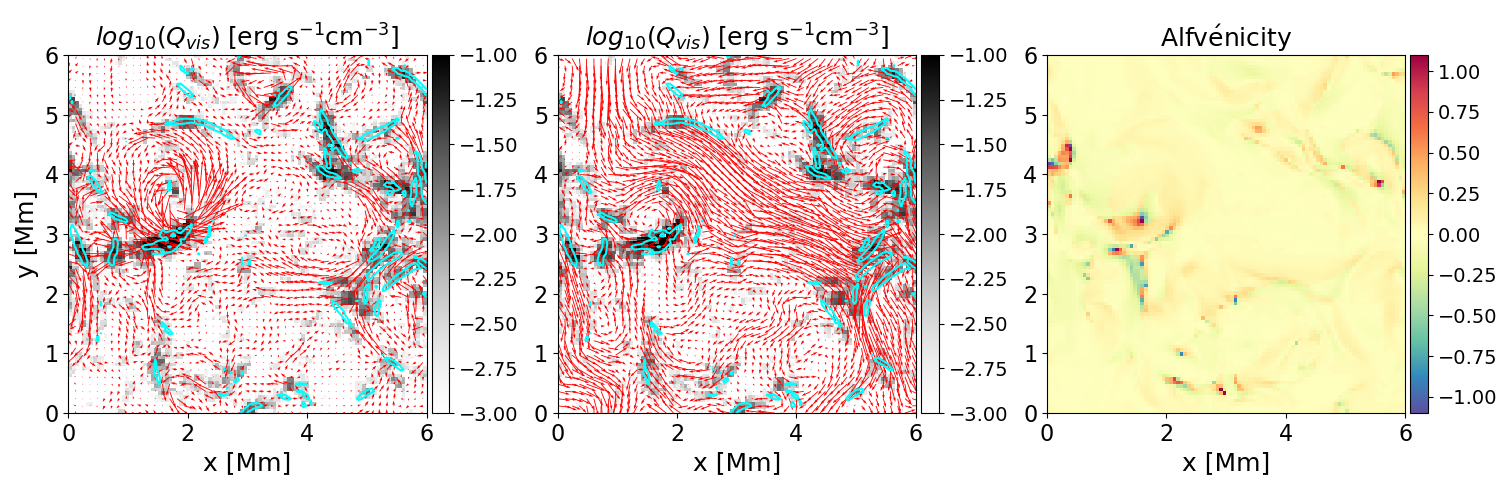}}
  \caption{Relationship of viscous heating, current sheets and flows for run LR for the same snapshot as Fig.~\ref{fig:connect_fp}. In the left and middle panels the greyscale image shows the volumetric viscous heating rate at the loop apex.  The blue contours outline the current sheets at the loop apex (the contours are the same as in Fig.~\ref{fig:connect_fp} and the red arrows illustrate the velocity field perpendicular to the loop axis (left panel), and the magnetic field perpendicular to the loop axis (middle panel). The right panel shows the Alfv\'enicity at the loop apex.}
  \label{fig:vel_qvis}
\end{figure*}
The viscous heating coincides well with the formation regions of current sheets. Determining whether the viscous heating is due to shear flows that generate the current sheets by shearing adjacent magnetic field lines or reconnection jets that thermalize is not trivial since a reconnection outflow from one event could be a reconnection inflow for another. In case of the jet scenario, the viscous heating should mostly occur at the ends of current sheets, while being enhanced along them in the shear scenario. The perpendicular velocity field and magnetic field overlayed with the positions of current sheets are depicted in Fig.  \ref{fig:vel_qvis}. Since the viscous heating is not clearly localized near the current sheet end points and most current sheets do not have clearly identifiable jets at their end points, the heating likely stems mostly from shear flows. This picture could change in simulations with higher resolutions that have a smaller viscous diffusivity. In addition, the smaller magnetic diffusivity would likely lead to current sheets with larger aspect ratios and consequently higher outflow speeds.\\

The twisted structure in the upper left quadrant is clearly discernible in both the perpendicular velocity and magnetic field components. The magnetic field direction has the opposite sign of the velocity, which is a property of Alfv\'en waves if the wave propagates in the direction of the magnetic guide field.  In an incompressible plasma in magnetohydrostatic equilibrium, the relation for magnetic field and velocity perturbations for an upward propagating torsional Alfv\'en wave is
\begin{equation}
    \mathbf{v}=-\frac{v_{A}}{B_{z}}\mathbf{b},
\end{equation}
with $\mathbf{v}$ being the velocity perturbation, $\mathbf{b}$ the magnetic field perturbation and $v_{A}$ the Alfv\'en speed. The ratio $q=-|B_{z}\mathbf{v}/(v_{A}\mathbf{b})|\frac{\mathbf{v}\cdot\mathbf{b}}{|v||b|}$ is shown in the third panel of Fig. \ref{fig:vel_qvis}. 
Here we use the velocity and magnetic field components perpendicular to the loop axis as a proxy for the velocity and magnetic field perturbations. The magnetic field perturbation is larger than expected from Alfv'en waves for most of the domain, suggesting that the perpendicular component of the field results from longer-lasting flows stretching the field lines. However, the ratio is enhanced within the prominent vortex flow, suggesting that this structure has both properties of torsional Alfv\'en waves and long-lived twisted structures.

\begin{figure}
\resizebox{\hsize}{!}{\includegraphics{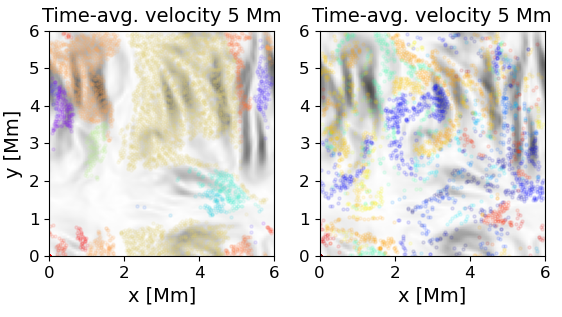}}
  \caption{Mapping of magnetic flux bundles and long-lived vortices for run LR. Left panel: Magnitude of the velocity vector computed from perpendicular velocity components averaged over a time window of five minutes at a distance of 5 Mm from the photosphere. The colors show the intersection of magnetic field lines with the slice at 5 Mm height and indicate in which magnetic concentration the respective field line is rooted. Right panel: Same as left panel, but with field line mapping to the opposite footpoint FP 2.}
  \label{fig:vel_magflux}
\end{figure}

Flux bundles originating from a magnetic source can rotate as a whole or be subdivided by vortex flows. Fig. \ref{fig:vel_magflux} shows the velocity field at a height of five Mm above the photosphere at FP 1. The coloured circles show the intersection of field lines with this plane, coloured according to the magnetic field concentration in which they are rooted. The vortex in the upper left quadrant comprises one bundle of field lines (orange) originating from the same patch at FP 1, suggesting that the structure rotates as a whole. In contrast to that, the adjacent larger structure contains multiple long-lived vortex and shear flows. Each bundle maps to multiple concentrations at FP 2, further complicating the identification of coherent structures. 
While long-lived vortices lead to further substructure within magnetic flux bundles, there is no one-to-one correspondence with the widths of observed coronal structures.



\section{Example of a reconnection event driven by a twisted flux rope}
\label{subsect:recon}

\begin{figure}
\resizebox{\hsize}{!}{\includegraphics{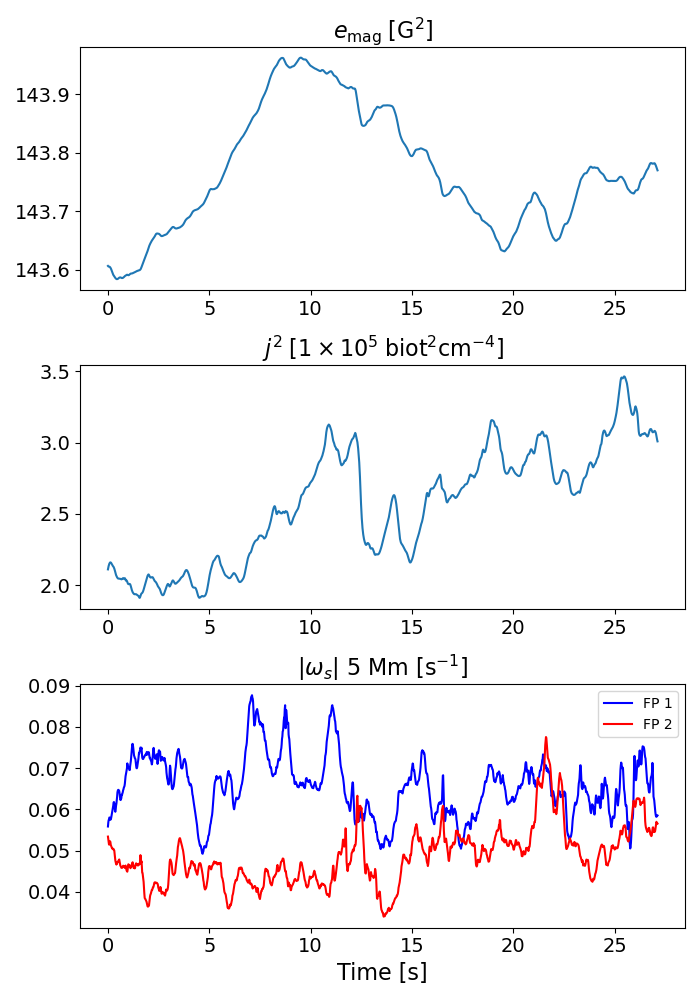}}
  \caption{Effect of a long-lived vortex flow on coronal quantities for run LR. Top to bottom: Magnetic energy density in the coronal part of the simulation domain averaged over a three-dimensional region with the extent $x\in [0,3]\; \rm{Mm}$, $y \in [2,5]\; \rm{Mm}$ and $s \in [5,45]\; \rm{Mm}$, squared current density averaged over the same region, averaged magnitude of the axial component of the vorticity at a height of 5 Mm above the photosphere for each footpoint. The vorticity was averaged over the 2D cross-section of the selected part of the simulation domain. Blue line: FP1, red line: FP2.}
  \label{fig:vort_time}
\end{figure}

\begin{figure*}
\resizebox{\hsize}{!}{\includegraphics{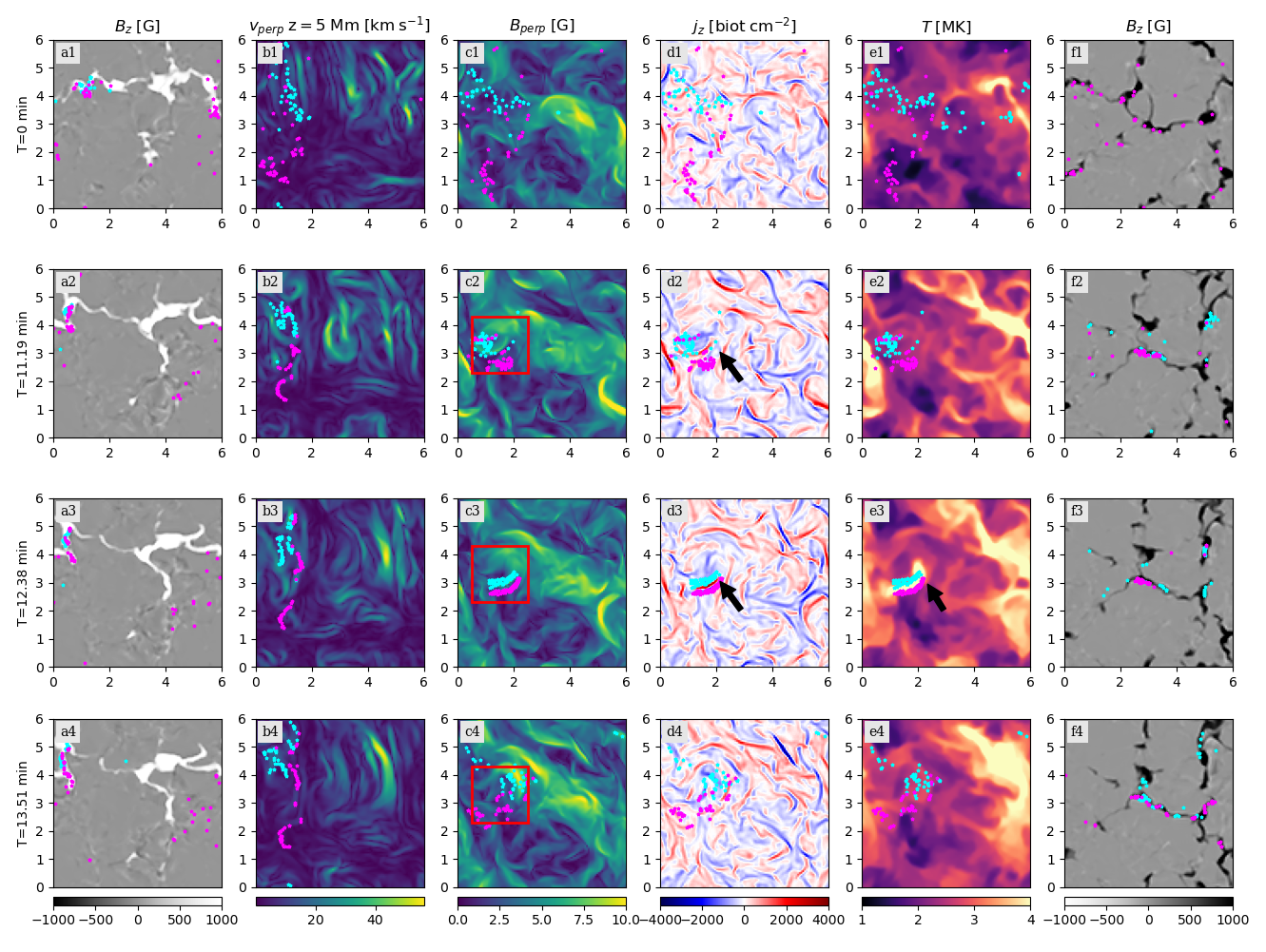}}
  \caption{Time sequence for a single reconnection event for run LR. From left to right column: Vertical component of the magnetic field $B_{z}$ at $z$=0 km (a), velocity perpendicular to the loop axis at a height of 5 Mm above the photosphere from where the magnetic field lines are tracked (b), magnetic field component perpendicular to the loop axis at the loop apex (c), loop-aligned current density at the loop apex (d), temperature at the loop apex (e) and vertical magnetic field component at $s$=50 Mm (f). The blue and pink stars show the intersection of the two traced populations of magnetic field lines with the respective plane. At $t$=39.98 min the two populations are aligned at each side of the current sheet.}
  \label{fig:recon}
\end{figure*}
Long-lived vortex flows can generate twisted flux ropes and therefore increase the magnetic energy in the corona, which can be released at a later time through reconnection.
The influence of a long-lived vortex flow on coronal magnetic energy is illustrated in Fig. \ref{fig:vort_time}. The magnetic energy density in the region of the simulation domain containing the vortex flow increases, reaches a peak shortly before the main reconnection event at 40 min and then decreases. The squared current density peaks at about 40 min, then rapidly falls off before increasing again. The vorticity averaged over a plane at $s=5\; \rm{Mm}$ above the photosphere shows several strong peaks before the reconnection event, while the vorticity at the opposite footpoint remains lower.\\

The persistent vortex flow can help create gradients in the velocity and magnetic field, drive magnetic field lines into the current sheet forming at its edge, and thus facilitate reconnection. The time evolution of magnetic field, velocity, current sheets and temperature for the reconnection event is shown in Fig. \ref{fig:recon}.\\
To investigate the influence of the vortex on reconnection and heating, we track magnetic field lines in time around the current sheet created by the vortex.
Fifty seed points for magnetic field lines were selected at the loop apex at time t=12.36 min on each side of the current sheet at the vortex edge (panel d3). The magnetic field lines are then evolved backwards and forwards in time using the method outlined in appendix~\ref{appsec:Btrack}. As stated previously, it is strictly speaking not possible to identify the same field line at different timesteps since the frozen-in condition is not fulfilled. However, the field line tracking should still allow us to obtain an idea of the evolution of the bundle of magnetic field lines as a whole.\\

Vortex flows are present during the entire duration of the event in the upper left quadrant of the loop cross-section at $s=5$ Mm (column b). The perpendicular magnetic field component is enhanced and the magnetic field forms a twisted structure (see the red boxes  in panels c2-c4 of Fig. \ref{fig:recon}) which is subsequently destroyed. A current sheet forms at the edge of the twisted structure and weakens after it is disrupted (black arrows in panel d2 and d3). A short-lived temperature enhancement to several million Kelvin is present at the location of the current sheet (see the black arrow in panel e3).\\

\begin{figure*}
\resizebox{\hsize}{!}{\includegraphics{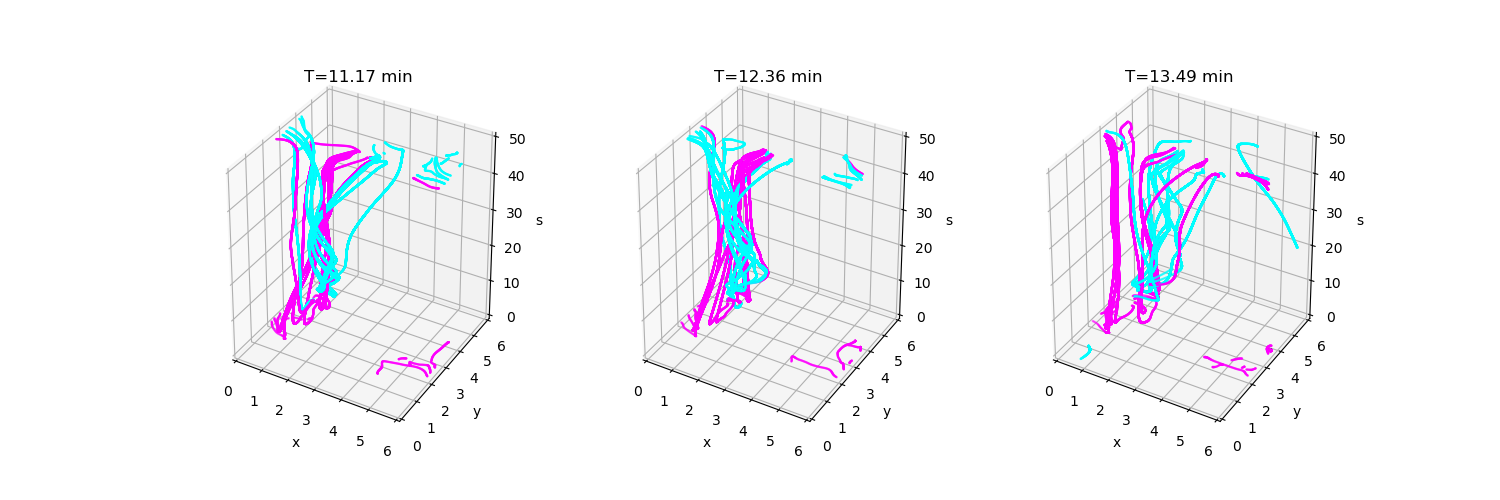}}
  \caption{3D rendering of magnetic field lines for three different times around the investigated reconnection event at $t$=12.36 min for run LR. The blue and pink colouring indicates the positions of the seedpoints relative to the current sheet studied in Sect. \ref{subsect:recon}.}
  \label{fig:recon3D1}
\end{figure*}

The magnetic field lines are concentrated by the vortex flow. At FP 1, field lines from either side of the current sheet are anchored close together, partially even in the same magnetic patch (column a). On the opposite footpoint, on the other hand, the field lines are distributed over multiple magnetic concentrations (column f).\\
To obtain an idea of the three-dimensional structure of the magnetic field around the reconnection event, Fig.~\ref{fig:recon3D1} shows 3D plots of the magnetic field lines shortly before, during and after the main reconnection event.
At the beginning of the event, the blue field lines have a twisted appearance. The structure consists mainly of two strands rooted in two different concentrations at FP 2 that are being twisted around each other at FP 1. The pink field lines are partially intertwined with the twisted structure formed by the blue field lines. During the reconnection event the blue field lines still appear twisted, while afterwards the twisted bundle is unravelled and the blue field lines connect to several magnetic patches at FP 2. 
While the current sheet forming at the edge of the twisted flux bundle is in some sense a tectonics current sheet, part of the field lines on each side of the sheet originate from the same magnetic patch, suggesting that internal braiding also plays a role in the reconnection.

\section{Discussion}
\label{sec:disc}

In this article, we have investigated the interplay between magnetic topology, heating events and observable structures for a coronal loop simulation that includes a self-consistent treatment of magnetoconvection and the coupling to the corona. We also compare it to idealized coronal heating models.

\subsection{Hybrid tectonics-braiding picture}
The preferred location of current sheet formation and heating differs for existing models of coronal heating.
In classical tectonics models, reconnection mostly occurs at the interfaces of sheared magnetic flux bundles.
Previous studies have confirmed this \citep[e.g.,][]{2002_Priest,2025_Howson}. In the original Parker braiding model, the field is initially uniform and there is no a-priori preferred site for heating. In our model, the field is concentrated in intergranular lanes, but field lines can be additionally tangled by internal motions within these magnetic patches. Due to these internal motions, reconnection between field lines can take place within one bundle but also in a more tectonics-like fashion between flux bundles connected to different magnetic concentrations. If internal braiding of magnetic field by motions within a magnetic flux concentration is the dominant heating mechanism, heating would occur throughout the cross-section of a magnetic flux bundle instead of being concentrated at its edges. Clearly distinguishing between the two is not possible since an "internal" reconnection with respect to a magnetic flux patch at one footpoint could correspond to a large jump of the reconnected fieldline between different flux concentrations at the other footpoint. Identifying two reconnecting "strands" is non-trivial since perturbations propagating in the corona driven by vortex and shear flows at the boundaries can reach much larger amplitudes than the translational speed of magnetic flux patches at the photosphere and enhance gradients in the magnetic field at the edges of flux bundles as well as internally, leading to heating. There is therefore no clear distinction between reconnection caused by "tectonics" or internal braiding.\\

Due to the low resistivity in the corona, most of the energy freed up by heating events is dissipated due to viscous heating. This, however, does not change the preferred heating locations in the different models, since
small scales in the velocity field also form preferentially at flux tube interfaces in the tectonics picture due to the relative motion of photospheric magnetic flux elements and hence coronal flux tubes.
In our simulation, which includes a shallow convection zone, each bundle of field lines originating from a magnetic patch at one footpoint loses its coherency along the loop, leading to a highly fragmented structure making it impossible to identify simple magnetic flux tubes in the classical sense.\\

Despite this complexity, we still attempt to determine whether current sheets preferentially form at the interfaces of magnetic flux coming from different concentrations taking into account the different mapping from both footpoints. In the low resolution case, about two-thirds of longer current sheets overlap with flux tube edges or interfaces. This is consistent with the findings of \citet{2018_Kanella}, who find the most elongated and largest heating events at interfaces of regions with different magnetic connectivity. Going to higher resolutions, however, this percentage decreases, indicating that internal braiding or a turbulent cascade to small scales becomes more important. For higher resolution, however, the fraction of "tectonics" current sheets increases with the threshold put on the current density, suggesting that the strongest current sheets still tend to form at interfaces between flux systems. Considering the fragmentation of flux bundles creating small scales on the same order as the sizes of internal flows, however, it cannot be clearly determined whether a current sheet results from relative motion between two flux systems or internal braiding. Nevertheless, it is still an interesting result that aspects of both tectonics and braiding are present and the magnetic field distribution at the photosphere does influence the preferred heating locations in the corona. The heating is thus not completely independent of the driving motions, as predicted by MHD turbulence models.\\
\subsection{Consequences for the interpretation of observations}
As a consequence of different heating sites, the expected locations of loop brightenings also differ between models. While heat conduction would in any case lead to the appearance of elongated strands, the interfaces of flux tubes would be expected to appear as bright strands in the tectonics model.\\

The factors determining the observed sizes of bright coronal loop strands are subject to speculation. 
In an estimate based on magnetic flux conservation, \citet{2015_Klimchuk} estimated that a coronal loop with a coronal field strength of 100 G and a diameter of 1500 km maps onto seven photospheric  kilogauss flux concentrations. Similarly, with SoHO data \citet{2003_Close} found photospheric magnetic sources were connected on average to five other sources.\\

In this study, we compared emitting regions with magnetic connectivity.
We find emitting regions with complex shapes.
The paradigm of a compact bright coronal loop strand as a thin magnetic flux tube with roughly circular cross-section has been challenged earlier, such as in \citet{2022_Malanushenko}. Their study, however, did not examine the magnetic connectivity of emitting regions.
We find that bright strands do not correspond to coherent magnetic flux bundles, nor are they clearly concentrated at the interfaces of flux systems.\\ 
The lack of correlation between regions with high density and kilogauss flux tubes can be explained by the timescales of cooling and chromospheric evaporation. The density distribution will follow the distribution of past heating events.
Bright strands are observed to be significantly larger than postulated current sheet dimensions in the solar corona. In our simulations, we also find that emitting regions are wider than current concentrations. This effect likely arises from the time evolution of current sheets. Current sheets evolve and deform, heated field lines reconnect and conduct heat down to different regions at the footpoints, thus broadening heated areas. Heating events also tend to cluster above strong magnetic field concentrations persisting for an extended amount of time, thus leading to hot regions encompassing multiple individual small heating events.

\subsection{Relationship between magnetic flux bundles and flows}
Vortex and shear flows produce additional substructure in magnetic flux bundles. 
\citet{2015_Peter} suggested that the loop substructure is determined by the interplay of thin kilogauss magnetic flux tubes anchored in the photosphere and vortex motions inside magnetic concentrations. 
Magnetic tornadoes can cause local brightenings in simulations attributed to reconnection \citep{2024_Kuniyoshi}. However, the emission depends on the interplay between heating, enhanced temperature, density and wavelength passbands used to observe the emission. 
\citet{2024_Kuniyoshi} used a low magnetic field strength of 10 G, leading to the formation of an isolated twisted magnetic tornado. The higher magnetic field strength in our simulation leads to a higher filling factor of magnetic concentrations in the photosphere. In areas above intense magnetic field concentrations large enough to contain several vortices, vortices will interact, leading to complex flow patterns \citep{2021_Battaglia}. While local brightenings at vortex edges could appear at coronal loop strands, vortices do not necessarily appear bright in the corona and there is no one-to-one correspondence between extended vortex tubes and bright strands \citep{2023_Breu}. Neither clearly identifiable magnetic field line bundles nor plasma vortices correspond to bright strands.\\

Internal flows can, however, enhance velocity shear and the perpendicular magnetic field component at the boundaries of flux bundles compared to shear solely due to the bulk motion of flux tubes. The hottest areas are present at the location of long-lived current sheets, which in turn tend to form in regions with long-lived shear flows (see bottom row of Fig. \ref{fig:vtavg}).
The heating at the edge of the rotating structure appears to be caused by the rotation of the flux tube rather than translational motion. With increasing numerical resolution, both plasma flows within the narrow intergranular lanes and the turbulent cascade to small scales are better resolved, leading to more expected heating across flux bundles.
\section{Conclusions}
\label{sec:conc}
Our simulations show a strong indication of a mixed coronal heating ecosystem even within an isolated coronal structure. Instead of ruling out some previous models, increasing the complexity of the numerical simulation leads to a synthesis of aspects of multiple previously reported models.\\

Each bundle of magnetic field lines originating from a photospheric source splits up into multiple strands connecting to several magnetic concentrations at the other loop footpoint, already indicating that the concept of a flux bundle is ambivalent. Motions within magnetic concentrations lead to further twisting and braiding of the field lines. The model therefore shows both aspects of tectonics and braiding models. While we do not find clearly identifiable magnetic flux tubes in the simulation, the connectivity nevertheless influences the preferential location for current sheet formation. The internal braiding, however, leads to a more diffuse distribution of the heating over the loop cross section than in models only considering the bulk motion of magnetic elements.\\

The velocity field shows fluctuations on a range of timescales. Indeed, chromospheric and coronal heating may involve a subtle interplay between reconnection and wave effects, as stressed recently by \cite{srivastava25}. While some coronal flows are short-lived perturbations, long-lived flows driven by photospheric motion play a role in the formation of current sheets. Long-lived vortex flows can generate twisted magnetic field structures, leading to further substructure of the loop and can act as inflows into reconnection regions as well as enhance gradients in the magnetic field.\\

The numerical resolution has been shown to influence the observable structure forward modelled from loop simulations. The resolution also has an influence on the heating mechanism. While at least for low-resolution simulations current sheets tend to form at the interfaces between flux systems, current sheets can form throughout the domain. If the numerical resolution is increased, a larger fraction of current sheets forms within magnetic flux bundles. The tectonics-like heating in low-resolution large-scale active region simulations may be  partially an artifact of lacking resolution. However, when increasing the resolution, a given flux bundle may be recategorised as a set of separate sub-bundles, between which tectonics heating may occur. While the balance between tectonics and braiding changes for higher resolution, the overall picture that a mixture between the two processes is present remains unchanged.\\
Generally, the interplay of heating and cooling timescales of the plasma with the complex magnetic topology means that the  observed structures are not always a good indicator for the 3D magnetic structuring.\\

Overall, we find a rich hybrid picture of coronal heating, showing aspects of both braiding on small scales and flux tube tectonics as well as wave motions. While the complex nature of coronal heating complicates the interpretation of observations, it synthesizes instead of excludes previous models.\\\\

\section*{Acknowledgements}

The research leading to these results has received funding from the UK Science and Technology Facilities Council (consolidated grant ST/W001195/1). We gratefully acknowledge the computational resources provided by the Cobra supercomputer system of the Max Planck Computing and Data Facility (MPCDF) in Garching,
Germany. 
DP gratefully acknowledges support through an Australian Research Council Discovery Project (DP210100709). 
IDM received funding from the Research Council of Norway through its Centres of Excellence scheme, project number 262622.

\section*{Data Availability}

Due to their size, the data from the numerical simulations and analysis presented
in this paper are available from the corresponding author upon
 request.



\bibliographystyle{mnras}
\bibliography{paper} 




\appendix

\begin{appendix}

\section{Flux bundle segmentation}
\label{appsec:segmentation}
We start by identifying magnetic concentrations (patches) at the $\langle\tau\rangle =1$ layer by applying a threshold of 1 kG to the vertical magnetic field component. We then trace magnetic field lines from each grid cell at the apex and compute the intersection of the field line with the $\langle\tau\rangle =1$ layer at each loop footpoint. The field lines are traced by solving the field line equation
\begin{equation}
    \frac{d\mathbf{r}}{ds}=\frac{\mathbf{B}}{B},
\end{equation}
where $\mathbf{B}$ is the magnetic field vector, $\mathbf{r}$ is the position vector in space and $s$ is the displacement along a magnetic field line.
To integrate the field line equation, we use a 4th-order Runge-Kutta scheme with adaptive stepsize. We employ trilinear interpolation for subgrid interpolation of the magnetic field.
If the field line intersects the photosphere within one of the previously identified kilogauss-strength concentrations, we assign a unique number, grouping the magnetic field lines according to the magnetic patch in which the field line is rooted at each footpoint. If the field line does not intersect the photosphere in a kilogauss-strength concentration, the assigned value is zero.
Since each field line has a footpoint in each photosphere, we obtain two different field line mappings. Field lines originating from one specific concentration in footpoint 1 can end up in different concentrations in footpoint 2 and vice versa.
\section{Automatic current sheet categorisation}
\label{appsec:current_auto}

\begin{figure}
\resizebox{\hsize}{!}{\includegraphics{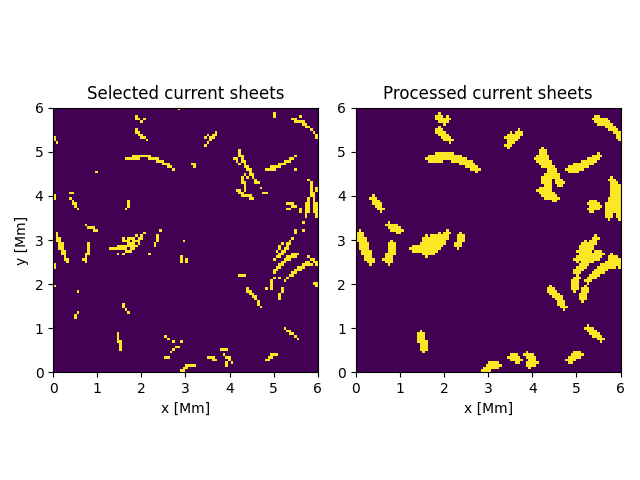}}
  \caption{Detection of current sheets for run LR and the same snapshot as Fig. \ref{fig:connect_fp}}. Left panel: Selected current sheets with current densities above the threshold. Right panel: Current sheets after removal of small structures and dilation of the area of the remaining current sheets.
  \label{fig:current_prep}
\end{figure}
\begin{figure*}
\resizebox{\hsize}{!}{\includegraphics{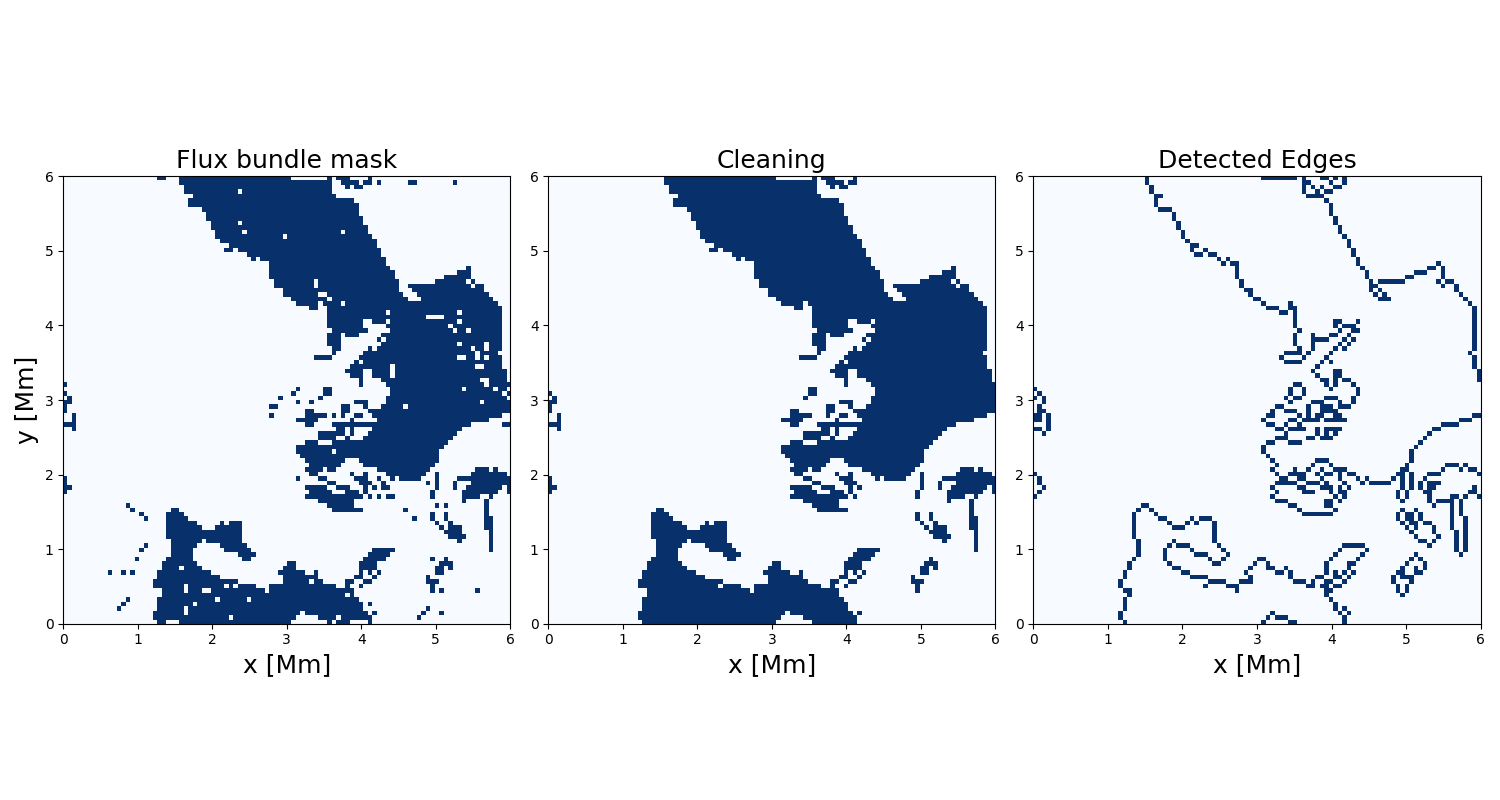}}
  \caption{Detection of magnetic flux bundle edges for run LR and the same snapshot as Fig. \ref{fig:connect_fp}. Left panel: Masked magnetic flux bundle. Middle panel: Cleaning by removing small structures and closing small holes, Right panel: Obtained mask outlining the edges by subtracting dilated and eroded mask.}
  \label{fig:edge_det}
\end{figure*}
A magnetic flux bundle is identified here as the grid cells threaded by magnetic field lines originating in the same magnetic concentration. Since field lines from a magnetic patch at one footpoint do not map one-to-one onto a magnetic patch at the other footpoint, we have two different possible maps of flux bundles at each height, one for each magnetic footpoint distribution.\\
We concentrate here on the relationship between current sheets and flux bundles at the loop apex. The identification and categorisation algorithm makes extensive use of python's open source image processing library scikit-image.
Current sheets are identified as 2D coherent regions at the loop apex with an unsigned current density above a threshold of $|j|\geq \langle |j|\rangle + a\sigma$ with $\sigma$ being the standard deviation of the magnitude of the current density. Both mean and standard deviation are computed over the loop cross section at the loop apex for each individual timestep. The identification of current sheets as connected structures in three dimensions poses additional challenges and is beyond the scope of this work.\\
Connected regions with a current density above the threshold are labelled using the function skimage.measure. Since current sheets form on the scales of the grid scale and often have widths of only one or two grid cells, we widen the structures by applying the dilation procedure from scikit.image \citep{scikit-image}. The result is illustrated in Fig. \ref{fig:current_prep}.\\
In order to categorise the current sheets, we test the overlap between current sheets and detected edges of flux bundles. To detect the edges of magnetic flux bundles, we first mask the area at the apex connected to a specific magnetic concentration at one footpoint. We close small holes in the mask with a size smaller than five grid cells and also remove small structures that cover less than five grid cells. To create a mask covering the edges of the resulting region, we first create two different masks by applying the dilation procedure from scikit.morphology. Dilation enlarges connected areas, by setting the value of a pixel to the maximum of all pixels in the local neighborhood. We now create a third mask by subtracting the dilated and original images. The result is a mask tracing the edges of magnetic flux bundles. This procedure is illustrated in Fig. \ref{fig:edge_det}. The algorithm then loops over each detected current sheet and multiplies the corresponding mask with the masks outlining the fluxtube edges, testing for overlap. The results are stored in a $M\times N$ matrix, with M being the number of current sheets and N the number of magnetic concentrations.
We repeat this procedure for both footpoints. 
\section{Field line tracking}
\label{appsec:Btrack}
Magnetic field lines are followed in time using a method similar to \citet{2021b_Yadav},
namely, assuming that the plasma and magnetic field evolve together due to the high conductivity of the solar corona. Since this assumption breaks down in reconnection regions of very high current density, care has to be taken regarding the interpretation of the results. Due to the numerical resistivity, there is no perfectly ideal region in the simulation box where the frozen-flux theorem is guaranteed to hold.
The timescale $\tau_{diff}$ for the slippage of the magnetic field through the plasma is 
\begin{equation}
   \tau_{diff} = \frac{L^{2}}{\eta},
   \label{eq:diff}
\end{equation}
where $L$ is the length scale of interest and $\eta$ is the magnetic diffusivity. A difficulty in estimating the diffusion timescale arises from the reliance on numerical diffusivity in the MURaM code.
With the value $\eta=1.8\times 10^{11}\; \rm{cm^{2}s^{-1}}$ estimated in \citet{Breu_2025}, we obtain an estimate of 200 s for diffusion across one grid-point (with $L=60\; \rm{km}$). While the slippage may well play  a negligible role on larger scales outside current sheets, 200 s is comparable to the timescales of interest for reconnection. Due to the complexity and the steep gradients in the magnetic field, an error of the size of one grid cell can lead to a quite different field line trajectory. This is a conceptual problem for resistive codes and cannot be overcome by using a different tracking method. While we do not trust the accuracy of the time evolution of a single field line, tracing a large number of field lines in time should still provide an idea of the overall evolution of the magnetic field on larger scales. Coherent coronal structures, such as the twisted magnetic field bundle considered earlier, have diameters of 1-2 Mm. If we set the length scale in Eq. \ref{eq:diff} to 1 Mm, the corresponding timescale is about 56000 sec. The time needed for the entire magnetic fieldline bundle to slip through the plasma is thus much larger than the simulation runtime.   
\section{Following a reconnection event}
\label{appsec:track_recon}
We place a total of 100 seed points around a reconnection region at the apex, 50 at each side of the current sheet we chose to follow. As a starting timestep we choose t=12.36 min.   
From the seedpoints, the field lines are traced in both directions, using the method outlined in appendix \ref{appsec:segmentation}. We then use the velocity vector at the seed point in order to advect the seed to the next timestep. The time update of the seed point position is also calculated using a fourth-order Runge-Kutta algorithm. The velocity vector is interpolated in time between snapshots using linear interpolation. We then trace magnetic field lines from the advected seed locations, which are not necessarily located at the tracking height anymore. The intersection of the new fieldline with the tracking plane is then chosen as the new seedpoint location that is then advected in the next timestep.
We experimented with several different tracking heights. Tracking at the average height of the photosphere proved difficult due to frequent reconnection events in the low-lying complex magnetic canopy. The numerical diffusion is very high in the chromospheric layer due to the strong magnetic field gradients. On the real Sun, the chromosphere is only partially ionized, allowing significant cross-field diffusion \citep[e.g.,][]{moreno22}.
We therefore chose a height of 5 Mm above the photosphere since at this height, the magnetic field has become predominantly vertical, and so the complexity and diffusion of the field is greatly reduced.
\end{appendix}



\bsp	
\label{lastpage}
\end{document}